\documentclass[lettersize,journal]{IEEEtran}

\usepackage{cite}
\usepackage{amsmath,amssymb,amsfonts,bm}
\usepackage{mathtools}
\usepackage{multirow}
\usepackage{bigints}
\usepackage{textcomp}
\usepackage{graphicx}
\usepackage{textcomp}
\usepackage{algorithmic}
\usepackage{soul}
\usepackage{dsfont}
\usepackage{algorithm}
\usepackage{array}
\usepackage{comment}
\usepackage{esvect}
\usepackage{subcaption}
\usepackage{xcolor}
\usepackage{subcaption}
\newcolumntype{M}[1]{>{\centering\arraybackslash}m{#1}}
\newcommand*{\QEDA}{\null\nobreak\hfill\ensuremath{\blacksquare}}%

\newcommand\myeqb{\mathrel{\stackrel{\makebox[0pt]{\mbox{\normalfont\scriptsize (b)}}}{=}}}

\usepackage{stfloats} 

\begin{document}

\title{ Deterministic and Statistical Analysis of the DoF of Continuous Linear Arrays in the Near Field}
%
%
%

\author{Athanasios G.~Kanatas,~\IEEEmembership{Senior Member,~IEEE,}  Harris K.~Armeniakos,~\IEEEmembership{Member,~IEEE,} \\
         Harpreet S. Dhillon,~\IEEEmembership{Fellow,~IEEE}, and Marco Di Renzo,~\IEEEmembership{Fellow,~IEEE}
         
\thanks{An earlier version of this paper has been presented in part at the IEEE International Conference on Communications (IEEE ICC 2024) \cite{ICC2024}.}
\thanks{A. G. Kanatas and H. K. Armeniakos are with the Department
of Digital Systems, University of Piraeus, 18534 Piraeus, Greece (e-mail:
\{kanatas, harmen\}@unipi.gr).}
\thanks{H. S. Dhillon is with with Wireless@Virginia Tech, Bradley Department of
Electrical and Computer Engineering, Virginia Tech, Blacksburg, 24061, VA, USA (e-mail: hdhillon@vt.edu).}
\thanks{M. Di Renzo is with Universit\'e Paris-Saclay, CNRS, CentraleSup\'elec, Laboratoire des Signaux et Syst\`emes, 3 Rue Joliot-Curie, 91192 Gif-sur-Yvette, France. (marco.di-renzo@universite-paris-saclay.fr), and with King's College London, Centre for Telecommunications Research -- Department of Engineering, WC2R 2LS London, United Kingdom (marco.di\_renzo@kcl.ac.uk)}
}
\maketitle

\begin{abstract}
This paper examines the number of communication modes, that is, the degrees of freedom (DoF) in a wireless line-of-sight channel comprising a small continuous linear intelligent antenna array in the near field of a large one. The framework allows for any orientations between the arrays and any positions in a two-dimensional space assuming that the transmitting array is placed at the origin. Therefore, apart from the length of the two continuous arrays, four key parameters determine the DoF and are hence considered in the analysis: the Cartesian coordinates of the center of the receiving array and two angles that model the rotation of each array around its center. The paper starts with the calculation of the \textit{deterministic} DoF for a generic geometric setting, which extends beyond the widely studied paraxial case. Subsequently, a stochastic geometry framework is proposed to study the \textit{statistical} DoF, as a first step towards the investigation of the system-level performance in near field networks. Numerical results applied to millimeter wave networks reveal the large number of DoF provided by near-field communications and unveil key system-level insights. A comparison of the proposed method with the singular value decomposition-based method is illustrated to validate the model.
\end{abstract}

\begin{IEEEkeywords}
Degrees of freedom, continuous linear arrays, large intelligent surfaces, near field, stochastic geometry.
\end{IEEEkeywords}

%
\IEEEpeerreviewmaketitle

\section{Introduction}
\IEEEPARstart{O}{ne} of the promising technologies for 6G communications is holographic multiple-input multiple-output (HMIMO) \cite{Huang20}, \cite{Gong24}. This technology consists of extreme electrically large and nearly continuous reconfigurable intelligent surfaces of finite size, which can be treated either as continuous-aperture phased (CAP) arrays of an infinite number of infinitesimal antennas \cite{Dardari21b}, or as spatially discrete arrays (SPD), also called reconfigurable holographic surfaces (RHS) \cite{Deng23}. These surfaces are capable of applying unprecedented wave transformations to impinging electromagnetic fields and are thus capable of controlling the radio environment in their proximity to some extent \cite{diRenzo20}.
The need to understand the achievable fundamental limits of these intelligent surfaces in terms of orthogonal spatial communication modes is well recognized by many authors \cite{Muharemovic08,Migliore08,Zhu24}. The number of these modes, termed spatial degrees of freedom (DoF), especially those that have a strong coupling intensity, provides the number of independent and effective data streams that can be supported simultaneously in a wireless system. The DoF and their coupling intensity are two fundamental performance indicators of electromagnetic signal and information theory \cite{France18}$- \hspace{-0.15cm}$\cite{Marco24}.

When two HMIMO devices communicate in far-field channels, the electromagnetic waves can be approximated as plane waves, and the system capacity depends on the richness of the scattering environment. In strong line-of-sight (LoS) channels, only one dominant path (i.e., with a strong coupling intensity) is available. In rich scattering environments, there are multiple strongly coupled communication modes (DoF), and appropriately designed orthogonal spatial beams form an optimal basis for the available spatial dimensions \cite{Sayeed02}, \cite{Tse05}, making beamspace MIMO communication possible \cite{Kalis08,Barousis11,Brady13,Vasileiou13}.
Nevertheless, when considering wave propagation in intricate scattering environments, such as urban settings, where the wavelength is significantly shorter than the typical size of the scatterers, the wave scattering process may exhibit chaotic ray dynamics. This phenomenon necessitates the use of a stochastic Green’s function \cite{Lin20}, which serves as the probabilistic solution to the wave equation in chaotic media. This approach is grounded in random matrix theory. Another approach is presented in \cite{PizzoSang22}, where the authors developed a Rayleigh fading model. In this model, the channel impulse response is a spatially stationary circularly symmetric complex Gaussian electromagnetic random field. The authors obtained a Fourier spectral representation of the random channel that accurately describes the impulse response only asymptotically, i.e., as the normalized array size approaches infinity.

\par The increasing interest in large intelligent surfaces (LIS) along with the relevance of higher frequency bands, such as the millimeter wave (mmWave)  spectrum in current 5G systems and the sub-terahertz (THz) spectrum in future 6G systems, constitutes a significant change in communication paradigm, in which far-field communication models are inaccurate and need to be replaced by near-field channel models \cite{Liu23,Cui23}. Depending on the carrier frequency and the size of the LIS, the near-field region between two LISs may extend from tens to several hundreds of meters. Consequently, unlike previous wireless generations, future LIS-aided 6G systems need to be designed based on near-field channel models where spherical waves replace the widely utilized plane waves \cite{Lu22}. This results in new opportunities for system design, such as the possibility of beam focusing in the near-field, in contrast to the conventional beam steering capability in far-field channels \cite{Zhang23,Bjornson21,Myers22}. 
In \cite{Hu18}, the authors derived the DoF when an LIS communicates with multiple single-antenna terminals in front of the LIS. The main result is that $2/\lambda$ terminals per meter can be spatially multiplexed for 1-D terminal deployments, while $\pi/\lambda^2$ terminals can be spatially multiplexed per square meter for 2-D and 3-D terminal deployments. The authors of \cite{Pizzo20} derived the DoF of a point-to-point HMIMO system with isotropic scattering. The main result is that the DoF is $2L_x/\lambda$ for 1-D linear arrays, $\pi L_xL_y/\lambda^2$ for 2-D planar arrays, and $2\pi L_xL_y/\lambda^2$ for 3-D volumetric arrays, where $L_x, L_y$ are the lengths of the arrays. Recently, the authors of \cite{Pizzo22} applied the multidimensional sampling theorem and Fourier theory to study the Nyquist sampling and the DoF of an electromagnetic field under arbitrary scattering conditions. The main result is that the number of DoF per unit area is equal to the number of Nyquist samples per square meter to be able to reconstruct the field within a specified accuracy. The authors of \cite{Ji23} used the Fourier plane-wave series expansion for electromagnetic fields to evaluate the number of additional DoF in the reactive near-field, which is determined by the evanescent waves.

The author of \cite{Miller00} introduced a general approach to evaluate the number of effective communication modes and their strength between two volumes. The solution is cast in terms of an eigenproblem obtained from the wave equation and Green's function. An interesting result is that a heuristic number of well-coupled modes (DoF) between two parallel communicating surfaces of area $A_T$ and $A_R$, located at a distance $r$, is equal to $A_T A_R/(\lambda^2 r^2)$, provided that the two surfaces are in the paraxial regime. Under the paraxial setting, to elaborate, $r$ is such that the Fresnel approximation can be applied, that is, only the first two terms of the Taylor series expansion of $r$ are retained, and the surface apertures are small compared to the propagation distance. This implies that the spherical wavefront is approximated by a parabola, leading to the term \textit{parabolic wavefront model} \cite{Do23}. Following a similar methodology, the authors of \cite{Pu15} derived a closed form expression for the DoF of LoS channels as a function of the orientations of the arrays, while still assuming the paraxial setting.

Naturally, analysis of the DoF requires, especially in free-space channels, careful consideration of the transmission distance, the size of the surfaces, and the relative geometry of the surfaces. Thus, when the paraxial approximation, which is often assumed in the literature, is not fulfilled, that is, the transmission distance is comparable with the sizes of the surfaces, the calculation of the DoF is not straightforward, and alternative methods are required. More specifically, besides the numerical solution of the eigenproblem formulated in, e.g., \cite{Miller00}, the DoF can be obtained by applying two general methods, which are known as (1) the spatial bandwidth \cite{Bucci87}-\cite{Franceschetti11} and (2) Landau's methods \cite{Landau75}-\cite{Ruiz23}, with the latter being highly coupled with the eigenproblem in \cite{Miller00}. Even though these methods are quite general and can be applied to general network scenarios, including non-paraxial settings, explicit analytical expressions for the DoF are difficult to obtain in general network deployments, e.g., in non-paraxial settings. Approximation approaches based on the spatial bandwidth and Landau's eigenproblem, which can be applied to non-paraxial settings, can be found in \cite{Dardari20} and \cite{Ruiz23}, respectively. The impact of spatial blocking on the DoF has recently been analyzed in \cite{Chen24}. A short overview based on \cite{Dardari21b} and applied to continuous linear arrays can be found in \cite{Marco23}. To overcome these limitations, some ad hoc methods have recently been proposed. For example, the authors of \cite{Dardari21a}, have provided closed-form solutions for the DoF between a linear LIS and a linear small intelligent surface (SIS), which is of practical interest since the considered setting models the link between a base station and a mobile terminal. While the approach in \cite{Dardari21a} is a key step towards understanding non-paraxial settings, a limited geometric setup is considered, and the solutions are restricted to specific conditions. 

In the present paper, we are mainly interested in linear continuous arrays. In this context, the authors of \cite{Ding22} have recently introduced an analytical approach to compute the DoF based on the spatial bandwidth method. A crucial assumption in \cite{Ding22} is that the considered linear arrays are composed of an infinite number of ideal point sources and that the transmitting linear array is characterized by an azimuth-independent radiation pattern. Under these assumptions, it is sufficient to parameterize the relative direction of the receiving linear array by using a single polar angle. This allows for a simplified analysis, since the transmitting and receiving linear arrays are always in the visibility region of one another. The same authors generalize the analysis in \cite{Ding24}, by introducing analytical asymptotic expressions for spatial bandwidth, highlighting the importance of the network geometry on the DoF.

Although DoF evaluation has been investigated in conventional geometric settings, especially in paraxial setup, a comprehensive performance analysis remains elusive, especially from a system-level (or statistical) point of view \cite{Schober}. Using fundamental results from stochastic geometry theory, new spatial models and channel statistics can be conceptualized to capture the peculiarities of near-field communications. This will allow for the characterization of location-dependent DoF in the receiver's association policy, with major applications in network performance analysis.

In this present work, a new analytical framework is introduced to calculate the number of DoFs of a communication link between two coplanar continuous linear arrays, a large and a small one, under LoS channel conditions. The proposed approach can be applied to paraxial and non-paraxial settings. To this end, the arrays are modeled as electromagnetic lines in free space, obeying the Huygens-Fresnel principle. The two arrays are assumed to radiate only in one of the two half-planes, in order to consider realistic deployments in which the transmitting array radiates only towards the intended receiving array. The transmitting array uses a focusing function to concentrate the energy in the direction of a focal point on the receiving array. The calculation of the DoF is based on the calculation of the number of orthogonal focusing functions that fit within the length of the receiving array. This approach allows us to account for near-field channels and non-paraxial settings. Therefore, the effective lengths of the two mutually visible arrays play an important role in the calculation of the DoF, since the mutual visibility of the arrays also determines the focusing functionality. Another factor that determines the DoF is the projected lengths of the two linear arrays onto the line that is perpendicular to the line that connects the centers of the two arrays. These projections are highly dependent on the positions and orientations of the arrays.
\par To elaborate, the contributions of this paper are summarized as follows:
\begin{itemize}
    \item  For a pair of linear arrays lying on the same plane, we provide a methodology to characterize the complete or partial visibility conditions between the arrays, without any restrictions on their relative positions and orientations. The considered unconstrained geometric setting, in which the lines can transmit only in a half hemisphere, can, in fact, lead to the lack of visibility or to a partial visibility condition. In the latter case, the effective length (contributing to the computation of the DoF) of the two arrays is reduced.
    \item We develop an algorithm to determine the visibility conditions between any pair of linear arrays and the corresponding geometric parameters to compute the DoF. The geometric parameters of interest include the point of intersection, if the prolongation of one array crosses the other array, and the corresponding effective lengths of the arrays that determine the DoF.
    \item We extend the methodology given in \cite{Dardari21a} by introducing a Taylor series expansion of the distance $r$ with one more term and derive the corresponding focusing function. Then the DoF are calculated for almost any geometric deployment of the arrays, considering the placement of the arrays almost anywhere on the horizontal plane and any orientation of the arrays. 
    \item We provide insights into the dependence of the number of DoF as a function of the geometrical parameters, that is, on the coordinates of the centers of the two arrays, their effective lengths, and the rotation angles. Comprehensive numerical results for a wide range of paraxial and non-paraxial settings are presented.
    \item We validate the proposed method to compute the number of DoF against the SVD numerical method, by considering free space channels.
    \item Finally, to the best of the authors' knowledge, this is the first work that computes the statistical distribution of the DoF for typical network deployments, under a stochastic geometry framework, providing the probability density function (PDF) and the cumulative distribution function (CDF) of the DoF. This analysis constitutes a first step towards the derivation of system-level insights on the DoF in near-field channels.
\end{itemize}

The remainder of this paper is organized as follows. In Section II, the system model is presented. In Section III, analytical geometrical conditions to determine the visibility between two arbitrarily deployed linear arrays and an algorithm to calculate the parameters of interest are introduced. In Section IV and Section V, frameworks for the deterministic and statistical analysis of the DoF are presented, respectively. Numerical results are illustrated in Section VI. Finally, Section VII concludes the paper.

\section{System Model}

\begin{figure}
\centering
  \includegraphics[width=0.44\textwidth]{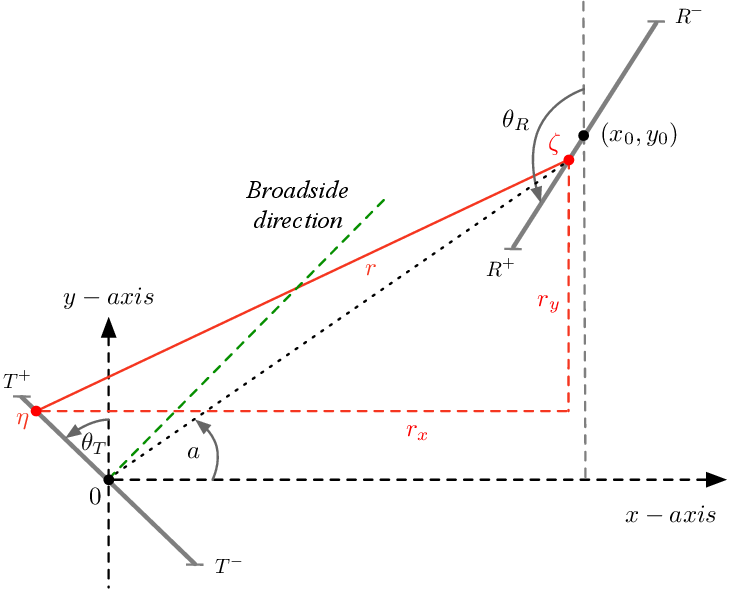}
  \caption{Considered system model depicting the capability of array rotation. The arrays appear to be of the same size to ease the depiction of the variables.}
  \label{fig_1}
\end{figure}

Consider two continuous linear antenna arrays as shown in Fig. \ref{fig_1}, where the origin of the reference system coincides with the center of the transmitting array, with the $x$ and $y$ axes oriented in the horizontal and vertical directions, respectively. The transmitting array, Tx, is a small-scale array. It has a length $L_T$ and rotates at an angle $\theta_T$ with respect to the $y$ axis, considering a positive angle for counterclockwise rotation and $-\pi < \theta_T\leq \pi$. The endpoints of the array are denoted as $T^+$ and $T^-$. The receiving array, Rx, is a large-scale array of length $L_R$, its center is located at $(x_0,y_0)$, and rotates by an angle $\theta_R$ with respect to the $y$ axis, considering a positive angle for counterclockwise rotation and $-\pi < \theta_R\leq \pi$. The endpoints of the Rx array are denoted as $R^+$ and $R^-$. Both arrays are assumed to be on the same plane. Moreover, we assume that both arrays radiate only in one of the two half-planes, in order to consider realistic deployments in which the transmitting arrays radiate only towards the intended receiving arrays. Let $\eta$, $-L_T/2\leq\eta\leq L_T/2$, denote the generic coordinate along the transmitting array and $\zeta$, $-L_R/2\leq\zeta\leq L_R/2$, denote the generic coordinate along the receiving array, assuming upward positive directions when $\theta_T=0^o$ and $\theta_R=0^o$, respectively. Thus, the coordinates of two generic points on the two arrays are $\left(\eta \sin\theta_T, \eta \cos\theta_T\right)$ and $\left(\zeta \sin\theta_R, \zeta \cos\theta_R\right)$, respectively. The distance $r$ between the two points $\eta$ and $\zeta$ on the two arrays is given by
\begin{equation} \label{dist}
\begin{aligned}
&r=\sqrt{r_x^2 + r_y^2} = \\
&\sqrt{(x_0 + \eta \sin \theta_T - \zeta \sin \theta_R)^2 + (y_0 - \eta \cos \theta_T + \zeta \cos \theta_R)^2}.
\end{aligned}
\end{equation}

\par
\begin{figure*}[!ht]
\centering
  \begin{subfigure}[t]{.34\linewidth}
  \includegraphics[trim=0 0 0 0,clip,width=\linewidth]{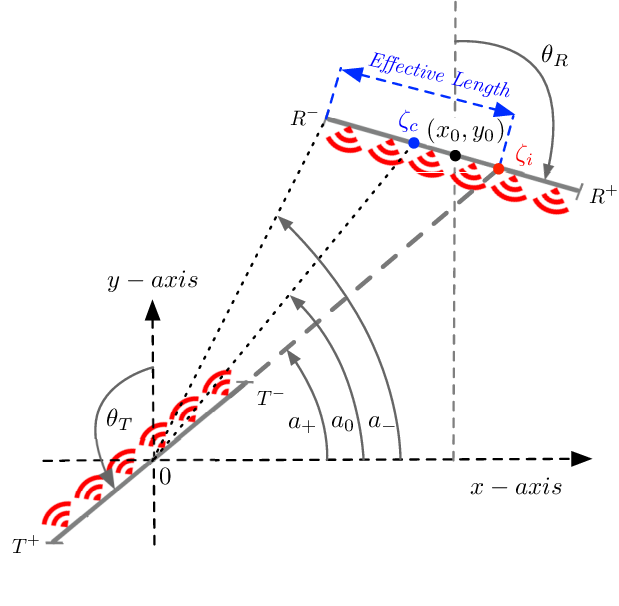}
   \caption{The Rx is intersected}
    \label{fig_2_a}
  \end{subfigure}\hfil
  \begin{subfigure}[t]{.34\linewidth}
    \includegraphics[trim=0 0 0 0,clip,width=\linewidth]{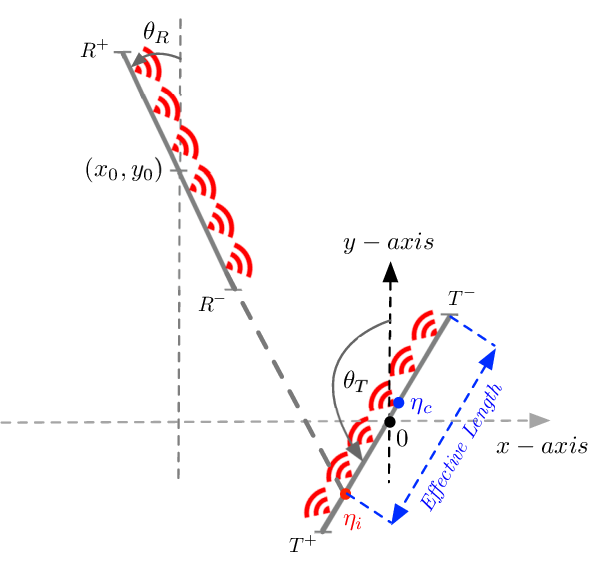}%
     \caption{The Tx is intersected}
    \label{fig_2_b}
  \end{subfigure}
    \caption{Two case studies with intersections}
  \label{Intersections}
\end{figure*}

 Since in the considered model the receiving array can be located anywhere on the plane and the two arrays may be rotated, there is the possibility that the receiving side of one array is not visible in its whole length from the transmitting side of the other array and vice versa. Two case studies with partial visibility are depicted in Fig. \ref{Intersections}. Therefore, based on the values of $x_0, y_0, \theta_T, \theta_R$, it is necessary to calculate the visible (effective) length of the two arrays $l_T, l_R$, as well as the points of intersection of the arrays, denoted as $\eta_i$ and $\zeta_i$, respectively. Based on the location of these points, one may calculate the new centers $-L_T/2\leq\eta_c\leq L_T/2$, $-L_R/2\leq\zeta_c\leq L_R/2$, of the two effective array segments, respectively, and replace $\eta$ with $ (\eta+\eta_c)$ and $\zeta$ with $(\zeta+\zeta_c)$ in (\ref{dist}), to account for the effective lengths. Moreover, $ (\eta-\eta_c) \in [-l_T/2,l_T/2]$ and $(\zeta-\zeta_c) \in [-l_R/2,l_R/2]$.

\par There are three additional angles to be defined for each array to facilitate the calculation of the DoF. These are the angles $a_+, a_-, a_0$, all defined in $(-\pi,\pi]$, measured from the center of the transmitter to the two endpoints $l_R/2$, $-l_R/2$, and the center of the receiver, respectively. The reference value, i.e., zero, is measured from the $x$ axis, as illustrated in Fig. \ref{fig_2_a}.
\par Based on a holographic assumption and a continuous phase profile as a function of $\eta$, one may define a focusing function at the Tx to focus the energy at the point $\zeta$ on the Rx, as
\begin{equation} \label{focus_fun}
F_T(\eta)|_\zeta = {\rm{rect}}\left(\frac{\eta}{l_T}\right) e^{j\frac{2\pi}{\lambda}r(\eta)},
\end{equation}
where $r(\eta)$ is defined in (1). If the Taylor series expansion is applied to $r(\eta)$ at $\eta=0$ and if one keeps the first three terms of the series, then
\begin{equation} \label{Taylor_1}
r(\eta) \approx r(0) + \eta \frac{\partial r(\eta)}{\partial \eta}\bigg\vert_{\eta=0} + \frac{\eta^2}{2} \frac{\partial^2 r(\eta)}{\partial \eta^2}\bigg\vert_{\eta=0}.
\end{equation}
In detail, the first- and second-order derivatives in (\ref{Taylor_1}) are given by (\ref{Deriv_1}) and (\ref{Deriv_2}), shown at the bottom of this page.
\begin{figure*}[h!b]
\hrulefill
\begin{equation} \label{Deriv_1}
    \frac{\partial r(\eta)}{\partial \eta} = 
    \frac{\eta + \eta_c - (\zeta + \zeta_c)\cos(\theta_R-\theta_T)-y_0\cos\theta_T + x_0\sin\theta_T}{\sqrt{[y_0 + (\zeta+\zeta_c)\cos\theta_R -(\eta+\eta_c)\cos\theta_T]^2 + [x_0 - (\zeta+\zeta_c)\sin\theta_R + (\eta+\eta_c)\sin\theta_T]^2}},
\end{equation}
\begin{equation} \label{Deriv_2}
    \frac{\partial^2 r(\eta)}{\partial \eta^2}
    = \frac{(x_0 \cos\theta_T - (\zeta+\zeta_c) \sin(\theta_R - \theta_T) + y_0 \sin \theta_T)^2}{[(y_0 + (\zeta+\zeta_c) \cos \theta_R - (\eta+\eta_c) \cos \theta_T)^2 + (x_0 - (\zeta+\zeta_c) \sin \theta_R + (\eta+\eta_c) \sin \theta_T)^2]^{3/2}}
\end{equation}
\end{figure*}
With the aid of some algebraic manipulations, it can be shown that the distance can be re-written as
\begin{equation} \label{Taylor_2}
r(\eta) \approx r(0) + \eta \, \rho + \eta^2 \, \tilde{\rho},
\end{equation}
where 
\begin{equation} \label{rho}
\rho = \frac{\sin \theta_T - \gamma \cos \theta_T}{\sqrt{1+\gamma^2}}, 
\end{equation}
\begin{equation} \label{rhotilde}
\begin{aligned}
\tilde{\rho} &= (x_0 - (\zeta + \zeta_c) \sin \theta_R + \eta_c \sin \theta_T)^{-\frac{5}{2}}\frac{(\cos \theta_T + \gamma \sin \theta_T)^2}{2(1+\gamma^2)^{3/2}} \\
&= (x_0 - (\zeta + \zeta_c) \sin \theta_R + \eta_c \sin \theta_T)^{-\frac{5}{2}} \frac{(1+\gamma^2)^{\frac{3}{2}}}{2} \left(\frac{\partial \rho}{\partial \gamma}\right)^2, 
\end{aligned}
\end{equation}
\begin{equation} \label{gamma_1}
\gamma = \frac{y_0+\zeta \cos \theta_R + \zeta_c \cos \theta_R- \eta_c \cos \theta_T}{x_0 - \zeta \sin \theta_R - \zeta_c \sin \theta_R + \eta_c \sin \theta_T} = \tan a, 
\end{equation}
and $a$ is the angle formed by the segment from the center of the Tx, $\eta_c$, to the point $\zeta$ and the $x$ axis (see Fig. \ref{fig_1}).
It is noted that the expansion in (\ref{Taylor_1}) is different from the expansion given in \cite{Do23} for the parabolic wavefront approximation, $r\approx r_x + \frac{r_y^2}{2d_0}$, where $d_0=\sqrt{x_0^2 + y_0^2}$, providing a considerably reduced approximation error in non-paraxial settings.
\par If one drops the terms that are independent of $\eta$, which contribute with a constant phase shift to the focusing function, then (\ref{focus_fun}) simplifies, using (\ref{Taylor_2}), to
\begin{equation} \label{focus_fun_2}
F_T(\eta)|_\zeta = {\rm{rect}}\left(\frac{\eta}{l_T}\right) e^{j\frac{2\pi}{\lambda}(\rho \, \eta + \eta^2 \tilde{\rho})}.
\end{equation}
Therefore, the approximation in (\ref{Taylor_2}) results in a quadratic phase as a function of $\eta$, which results in a focusing function in the near field. The requirement for a quadratic polynomial originates from the fact that the boresight Fraunhofer distance for a Tx linear array of length $L_T$ and an Rx linear array of length $L_R$ is $d^{ff}=\frac{2(L_T+L_R)^2}{\lambda}$, \cite{Cui23}, \cite{Monemi}.
Therefore, the Tx concentrates energy in the direction of the focal point on the Rx array. The distance of the two arrays also determines the depth of the focusing beam. If the Rx array is placed in the far field, then the focusing function degenerates into a beam steering phase profile, resulting in an infinite depth beamforming. Because of the reciprocity of the radio link, the DoF is the same if one switches the role of the two arrays from Tx/Rx to Rx/Tx. Therefore, there is no need to examine separately uplink and downlink transmissions, as far as the DoF is considered, as long as the definition of the orthonormal basis functions remains consistent.

\section{Geometric Conditions For Visibility}
The first geometric condition to be examined refers to the mutual visibility of the two arrays. As already explained, the Rx may be located anywhere on the plane. In practice, however, its center position, $(x_0,y_0)$, may be in a finite spatial region, e.g., a circular disk of radius $R$. This spatial region has an impact on the visibility of the two arrays. In addition, the two arrays may be rotated and this affects their mutual visibility. The rotation is governed by the angles $\theta_T,\theta_R$. In general, depending upon the location of the Rx, four different cases exist: i) no visibility, ii) full visibility, iii) the whole length of the Tx is visible to the Rx, but the prolongation of the transmitting linear array intersects the receiving linear array, as shown in Fig. \ref{fig_2_a}, and iv) the whole length of the Rx is visible to the Tx and the prolongation of the receiving linear array intersects the transmitting linear array, as shown in Fig. \ref{fig_2_b}. The analysis of the last two cases requires the calculation of effective length of the receiving or transmitting arrays, as illustrated in Fig. 2.

\subsection{Description of the Visibility Algorithm}
This subsection presents the main methodological steps to identify the visibility condition between the two arrays, which corresponds to identifying the case, among the four discussed, to be considered, as well as to compute the parameters required for the calculation of the number of DoF. The visibility condition is governed by the length and orientation of the two arrays. For this reason, two segments are defined using the endpoints of the arrays and two lines in parametric form.
Based on Fig. 1, the endpoints of the two arrays are given by
\begin{equation}
    \begin{aligned}
    T^+ & = \bigg(-\frac{L_T}{2}\sin \theta_T, \frac{L_T}{2}\cos \theta_T\bigg) \\
    T^- & = \bigg(\frac{L_T}{2}\sin \theta_T, -\frac{L_T}{2}\cos \theta_T\bigg), 
    \end{aligned}
\end{equation}
\begin{equation}
    \begin{aligned}
    R^+ & = \bigg(x_0-\frac{L_R}{2}\sin \theta_R, y_0 + \frac{L_R}{2}\cos \theta_R\bigg) \\
    R^- & = \bigg(x_0 + \frac{L_R}{2}\sin \theta_R, y_0 -\frac{L_R}{2}\cos \theta_R\bigg).
    \end{aligned}
\end{equation}
In addition, the center points of the transmitting and receiving matrices are denoted by $T^0=(0,0), R^0=(x_0,y_0)$, respectively. Based on the end points in (9) and (10), we introduce two functions that represent, in a parametric form, two lines between them, as follows:
\begin{equation}\label{Lines}
    \begin{aligned}
    t: \mathbb{R}\rightarrow\mathbb{R}^2 : \beta \mapsto \beta T^+ + (1-\beta) T^-, \\
    r: \mathbb{R}\rightarrow\mathbb{R}^2 : \delta \mapsto \delta R^+ + (1-\delta) R^-. \\
    \end{aligned}
\end{equation}
The image of the functions in (\ref{Lines}) are two lines between $T^+$ and $T^-$, and $R^+$ and $R^-$ respectively. If the domain, i.e., the parameters $\beta$ and $\delta$ are restricted to the interval $[0,1]$, the image is the segment between the endpoints. More precisely, if $\beta=0$, we obtain the point $T^-$, if $\beta=1$, we obtain the point $T^+$, and if $\beta=\frac{1}{2}$, we obtain the midpoint of the segment. The values $\beta<0$ correspond to the points of the line that extend beyond $T^-$, and the values $\beta>1$ correspond to the points that extend beyond $T^+$. The same applies to the parameter $\delta$. 
The two lines may intersect, and in such a case, the point of intersection needs to be calculated and subsequently verified to determine whether it falls within the Tx or Rx array. Additionally, the algorithm applies a line-side check, which entails determining whether the Rx array is positioned to the left or right of the Tx and, further, whether the receiving side of the Rx array faces the transmitting side of the Tx array. For these computations, a computer graphics approach is employed and additional vector definitions are introduced next, including the outer product for two-dimensional vectors.
\par Next, following a common computer graphics notation \cite{Hughes}, we define three vectors as the {\it{difference}} between points 
\begin{equation}
    \begin{aligned}
    \mathbf{t} & = (T^- - T^+) = [L_T \sin \theta_T \,\,\,\,\, -L_T \cos \theta_T]^T, \\
    \mathbf{r} & = (R^- - R^+) = [L_R \sin \theta_R \,\,\,\,\, -L_R \cos \theta_R]^T, \\
    \mathbf{c} & = (R^0 - T^0) = [x_0 \,\,\,\,\, y_0]^T,
    \end{aligned}
\end{equation}
which will be used for a line-side check, especially when the Tx lies in the receiving side of the Rx and/or the Rx lies in the transmitting side of the Tx. 
To this end, we introduce the magnitude of the cross product between two dimensional vectors as the determinant of the following $2\times2$ matrix:
\begin{equation}
\begin{aligned}
\mathbf{r}\times\mathbf{c} & \triangleq \left|\begin{array}{cc}
    r_x &  r_y \\
    c_x & c_y
\end{array}\right|    = r_xc_y-r_yc_x.
\end{aligned}
\end{equation}
The first line-side check is performed for the point $T^0$, based on the cross product 
\begin{equation}\label{cross_r_mc}
\mathbf{r}\times(-\mathbf{c}) = - L_R (y_0 \sin \theta_R +x_0 \cos \theta_R).
\end{equation}
Specifically, if $\mathbf{r} \times (-\mathbf{c}) > 0$, $T^0$ is located to the left of Rx and within its receiving half plane; if $\mathbf{r} \times (-\mathbf{c}) < 0$ then $T^0$ is located to the right of the Rx and on the other half plane; and if $\mathbf{r} \times (-\mathbf{c}) = 0$, $T^0$ is located on the Rx array. 
The second line-side check is performed for the $R^0$ point with respect to the Tx, by using the cross-product
\begin{equation}\label{cross_t_c}
\mathbf{t}\times\mathbf{c} = L_T(x_0\cos \theta_T + y_0 \sin \theta_T)
\end{equation}
and similar conclusions for $T^0$ can be drawn.
The vectors normal to $\mathbf{t}$ and $\mathbf{r}$ that are directed toward the transmitting and receiving side of the two arrays, respectively, are called {\it{inward edge normal}} vectors and are calculated using the cross product of two dimensional vectors  
\begin{equation}
    \begin{aligned}
    \mathbf{t}_n \triangleq \times \mathbf{t} \triangleq [-t_y \,\,\,\,\, t_x]^T = [L_T \cos \theta_T \,\,\,\,\, L_T \sin \theta_T]^T, \\
    \mathbf{r}_n \triangleq \times \mathbf{r} \triangleq [-r_y \,\,\,\,\, r_x]^T = [L_R \cos \theta_R \,\,\,\,\, L_R \sin \theta_R]^T.
    \end{aligned}
\end{equation}
These {\it{inward}} normal vectors are used to formulate the point of intersection between the two lines defined in (\ref{Lines}). Specifically, if $P$ is the point of intersection, then the parameters $\beta_P$ and $\delta_P$ that correspond to the point of intersection $P$ are calculated as follows:
\begin{equation}
    \begin{aligned}
    &\beta_P T^+ + (1-\beta_P) T^- = \delta_P R^+ + (1-\delta_P) R^- \\
    &\Rightarrow (T^- - R^-) = \beta_P \mathbf{t} - \delta_P \mathbf{r} \\
    & \Rightarrow (T^- - R^-) \cdot \mathbf{t}_n = - \delta_P (\mathbf{r} \cdot \mathbf{t}_n) \\
    & \Rightarrow \delta_P = - \frac{(T^- - R^-) \cdot \mathbf{t}_n}{\mathbf{r} \cdot \mathbf{t}_n} = \frac{1}{2} - \frac{x_0 \cos \theta_T + y_0 \sin \theta_T}{L_R \sin (\theta_T-\theta_R)},
    \end{aligned}
\end{equation}
and in a similar manner
\begin{equation}
    \beta_P = \frac{(T^- - R^-) \cdot \mathbf{r}_n}{\mathbf{t} \cdot \mathbf{r}_n} = \frac{1}{2} - \frac{x_0 \cos \theta_R + y_0 \sin \theta_R}{L_T \sin (\theta_T-\theta_R)}.
\end{equation}
The coordinates $(P_x,P_y)$ of the point $P$, given by (\ref{Pcoord}), shown at the bottom of the page, are calculated by substituting the solution for the parameter $\delta_P$ for the line $r$ or the parameter $\beta_P$ for the line $t$. The parameter $\zeta_i$ (the distance between $P$ and $R^0$) on the line $r$ of the Rx array is given by
\begin{equation} \label{intersectionpoint}
         \zeta_i = \pm \sqrt{(P_x - R^0_x)^2 + (P_y - R^0_y)^2}
         = \frac{x_0 \cos \theta_T + y_0 \sin \theta_T}{\sin (\theta_T-\theta_R)}
\end{equation}
where the plus sign is applied if $\delta_P > \frac{1}{2}$, and the minus sign is applied if $\delta_P < \frac{1}{2}$. In a similar manner, one may calculate the parameter $\eta_i$ on the line $t$ of the Tx 
\begin{equation}
         \eta_i = \pm \sqrt{(P_x - T^0_x)^2 + (P_y - T^0_y)^2}
         = \frac{x_0 \cos \theta_R + y_0 \sin \theta_R}{\sin (\theta_T-\theta_R)}.
\end{equation}
\begin{figure*}[!hb]
\hrulefill
\begin{equation}\label{Pcoord}
\begin{aligned}
    P&=\bigg(x_0+\frac{x_0 \cos \theta_T + y_0 \sin \theta_T}{\sin (\theta_T-\theta_R)} \sin \theta_R, \,\, y_0-\frac{x_0 \cos \theta_T + y_0 \sin \theta_T}{\sin (\theta_T-\theta_R)} \cos \theta_R\bigg)\\
    &= \bigg(\frac{x_0 \cos \theta_R + y_0 \sin \theta_R}{\sin (\theta_T-\theta_R)} \sin \theta_T, \,\, -\frac{x_0 \cos \theta_R + y_0 \sin \theta_R}{\sin (\theta_T-\theta_R)} \cos \theta_T\bigg),
\end{aligned}
\end{equation} 
\end{figure*}
Therefore, the new centers of the Tx and Rx arrays can be calculated when the point of intersection lies on their line segments, i.e., when $0 \leq \beta_P \leq 1$ or $0 \leq \delta_P \leq 1$, and the parameters $\zeta_i$ and $\eta_i$ take values in the range $[-L_R/2,L_R/2]$,  $[-L_T/2, L_T/2]$, respectively. Therefore, in partial visibility conditions, to calculate the effective lengths of the arrays, one has to differentiate between four cases: i) partial Tx visibility and the point $T^-$ is visible from the Rx array, ii) partial Tx visibility and the point $T^+$ is visible, iii) partial Rx visibility and the point $R^-$ is visible from the Tx array, and iv) partial Rx visibility and the point $R^+$ is visible. This is accomplished by defining four auxiliary vectors: $\mathbf{t_-} = (T^- -R^0)$, $\mathbf{t_+} = (T^+ -R^0)$, $\mathbf{r_-} = (R^- -T^0)$, and $\mathbf{r_-} = R^- -T^0$, and the following magnitudes of cross products:
\begin{equation}
\begin{aligned}
\mathbf{r}\times \mathbf{t^-} = & -\frac{L_R L_T}{2} \sin \theta_R \cos \theta_T - y_0 L_R \sin \theta_R  \\
&+\frac{L_R L_T}{2} \cos \theta_R \sin \theta_T - x_0 L_R \cos \theta_R,
\end{aligned}
\end{equation}
\begin{equation}
\begin{aligned}
\mathbf{r}\times \mathbf{t^+} = & \frac{L_R L_T}{2} \sin \theta_R \cos \theta_T - y_0 L_R \sin \theta_R  \\
&-\frac{L_R L_T}{2} \cos \theta_R \sin \theta_T - x_0 L_R \cos \theta_R,
\end{aligned}
\end{equation}
\begin{equation}
\begin{aligned}
\mathbf{t}\times \mathbf{r^-} = & -\frac{L_R L_T}{2} \cos \theta_R \sin \theta_T + y_0 L_T \sin \theta_T,  \\
&+\frac{L_R L_T}{2} \sin \theta_R \cos \theta_T + x_0 L_T \cos \theta_T
\end{aligned}
\end{equation}
\begin{equation}
\begin{aligned}
\mathbf{t}\times \mathbf{r^+} = & \frac{L_R L_T}{2} \cos \theta_R \sin \theta_T + y_0 L_T \sin \theta_T  \\
&-\frac{L_R L_T}{2} \sin \theta_R \cos \theta_T + x_0 L_T \cos \theta_T.
\end{aligned}
\end{equation}. 
The coordinates of the new center on the Rx array when the endpoint $R^-$ is visible from the Tx array are given by
\begin{equation}\label{AntCentre}
    C^R = \bigg(\frac{1}{2}P_x + \frac{1}{2}R^-_x, \,\,  \frac{1}{2}P_y + \frac{1}{2}R^-_y \bigg).
\end{equation}
When $R^+$ is visible from the Tx array, it is necessary to replace $R^-$ with $R^+$ in (\ref{AntCentre}). Then, the distance between $C^R$ and $R^0$ is 
\begin{equation}\label{Rxcentre}
    \zeta_c = \left\{ \begin{array}{rcl} \frac{\zeta_i}{2}+\frac{L_R}{4} &\mbox{if} & R^- \; \mbox{is visible}\\
   \frac{\zeta_i}{2}-\frac{L_R}{4} &\mbox{if} & R^+ \; \mbox{is visible}
    \end{array}\right.
\end{equation}
Following the same procedure for the Tx array, the distance of the new center from the origin is given by replacing $L_R$ with $L_T$, $\zeta_i$ with $\eta_i$, and $R^-,R^+$ with $T^-, T^+$ in (\ref{Rxcentre}). 
Finally, the new lengths of the arrays are given by
\begin{equation}\label{Rxlength}
    l_R = \left\{ \begin{array}{rcl} \big|\zeta_i -\frac{L_R}{2}\big| &\mbox{if} & R^- \,\, \mbox{is visible}\\
    \big|\zeta_i + \frac{L_R}{2}\big| &\mbox{if} & R^+ \,\, \mbox{is visible},
    \end{array}\right.
\end{equation}
and
\begin{equation}\label{Txlength}
    l_T = \left\{ \begin{array}{rcl} \big|\eta_i -\frac{L_T}{2}\big| &\mbox{if} & T^- \,\, \mbox{is visible}\\
    \big|\eta_i + \frac{L_T}{2}\big| &\mbox{if} & T^+ \,\, \mbox{is visible}.
    \end{array}\right.
\end{equation}
\subsection{Proposed Visibility Algorithm}
Based on the structure and definitions presented in the preceding subsection, the algorithmic steps to assess the visibility condition between the two arrays and to compute the new centers and effective lengths are provided in Algorithm 1.
\begin{algorithm}
\caption{Check Visibility Status for DoF Calculation}
\begin{algorithmic}\label{Algo_Vis}
\STATE Calculate $\beta_P$ and $\delta_P$
\IF{$\beta_P \in [0,1]$ \AND $\delta_P \in [0,1]$}
\STATE $DoF \gets NaN$ \COMMENT{The Tx touches the Rx}
\ELSIF{($\beta_P > 1$ \OR $\beta_P < 0$) \AND ($\delta_P > 1$ \OR $\delta_P < 0$)}
\IF{$\mathbf{t}\times\mathbf{c} > 0$ \AND $\mathbf{r}\times (-\mathbf{c}) > 0$}
\STATE Calculate $DoF$ with full length \COMMENT{Full Visibility}
\ELSE
\STATE $DoF \gets 0$ \COMMENT{No Visibility}
\ENDIF
\ELSIF{$\beta_P > 0$ \AND $\beta_P < 1$}
\IF{$\mathbf{t}\times\mathbf{c} > 0$ \AND $\mathbf{r}\times (\mathbf{t_-}) > 0$}
\STATE Partial Tx Visibility. Point $T^-$ is visible
\STATE Calculate $l_T$, $\eta_c$, and $DoF$
\ELSIF{$\mathbf{t}\times\mathbf{c} > 0$ \AND $\mathbf{r}\times (\mathbf{t_+}) > 0$}
\STATE Partial Tx Visibility. Point $T^+$ is visible
\STATE Calculate $l_T$, $\eta_c$, and $DoF$
\ENDIF
\ELSIF{$\delta_P > 0$ \AND $\delta_P < 1$}
\IF{$\mathbf{r}\times (-\mathbf{c}) > 0$ \AND $\mathbf{t}\times (\mathbf{r_-}) > 0$}
\STATE Partial Rx Visibility. Point $R^-$ is visible
\STATE Calculate $l_R$, $\zeta_c$, and $DoF$
\ELSIF{$\mathbf{r}\times (-\mathbf{c}) > 0$ \AND $\mathbf{t}\times (\mathbf{r_+}) > 0$}
\STATE Partial Rx Visibility. Point $R^+$ is visible
\STATE Calculate $l_R$, $\zeta_c$, and $DoF$
\ENDIF
\ENDIF
\end{algorithmic}
\end{algorithm}
\section{Calculation of the DoF}
The methodology used for the calculation of the DoF is based on the eigenfunction problem initially proposed in \cite{Miller00} for optical systems and then in \cite{Dardari20} for RF systems and is based on Green's function. The Green's function is equivalent to the spatial channel impulse response between any two points of the Tx and Rx arrays. It relates the transmitter’s current density distribution, $\phi(\eta)$, and the receiver’s electric field, $\psi(\zeta)$, via the following spatial integral:
\begin{equation}
    \psi(\zeta) = \int_{-l_T/2}^{l_T/2} G(\zeta,\eta) \phi(\eta) {\rm{d}} \eta.
\end{equation}
Each array is considered as a continuum composed of an infinite number of infinitesimal antennas, each producing a spherical wave, i.e., it is a Huygens' source. Although the transmitted field is a vector, it is approximated as a complex scalar to simplify the analysis. Correspondingly, in this work, Green's function is not a tensor but a scalar function. Polarization effects may be incorporated by replacing every antenna point by three mutually perpendicular electric dipoles \cite{Poon05}. The evaluation of the DoF consists of decomposing the spatial channel into a series of independent parallel sub-channels by determining the equivalent “singular values” through an eigenvalue decomposition of the Hermitian kernel of Green’s function. The resulting eigenfunctions constitute two complete sets of orthogonal basis functions, one associated with the Tx array and the other with the Rx array. The number of non-zero eigenvalues of Green’s function kernel, is defined as the number of DoF. However, the analytical solution to this problem for a generic geometric setup is highly challenging. An alternative solution was proposed in \cite{Dardari21a} using the kernel functions. The kernel function for two points $\zeta, \zeta'$ on the Rx array is given by
\begin{equation} \label{kernel_1}
K_R(\zeta,\zeta') = \int_{-\frac{l_T}{2}}^{\frac{l_T}{2}} G(\zeta,\eta)G^*(\zeta',\eta) {\rm{d}\eta}
= \int_{-\frac{l_T}{2}}^{\frac{l_T}{2}} \frac{e^{-jk(r-r')}}{(4\pi)^2 \, rr'} \rm{d}\eta
\end{equation}
The distances $r$ and $r'$ in the denominators can be approximated by the distance between the centers of the two arrays $d_0=\sqrt{x_0^2 + y_0^2}$. As explained in \cite{Sherman62}, if the distance $d_0$ between the two arrays satisfies $d_0 \geq 1.2 \, (L_T + L_R)$, the variations of the amplitude are negligible and thus the approximation $r\approx r' \approx d_0$ for the amplitude is valid. This approximation cannot be adopted for the phases in (\ref{kernel_1}).
The kernel function can be written with the help of the focusing function defined in \eqref{focus_fun_2}, as follows:
\begin{equation} \label{kernel_2}
K_R(\zeta,\zeta') \approx \frac{1}{(4\pi d_0)^2}\int_{-l_T/2}^{l_T/2} e^{-j\frac{2\pi}{\lambda}(\rho -\rho') \eta} e^{-j\frac{2\pi}{\lambda}(\tilde{\rho} -\tilde{\rho}') \eta^2} \rm{d}\eta
\end{equation}
This kernel is identical to the field distribution at the Rx array when the phase profile at the Tx array is set to focus towards $\zeta'$ 
\begin{equation}
    \psi(\zeta)|_{\zeta'} = \int_{-l_T/2}^{l_T/2} G(\zeta,\eta) F_T(\eta)|_{\zeta'} {\rm{d}} \eta.
\end{equation}
Using \cite[eq. 2.33.3]{Ryzhik} the kernel function in (\ref{kernel_2}) is obtained in closed form in (\ref{int_erf}), shown at the bottom of the next page, where $\rm{Erfi}(\cdot)$ denotes the imaginary error function.

\begin{figure*}[!hb]
\hrulefill
\begin{equation}\label{int_erf}
 K_R(\zeta,\zeta') \approx \frac{1}{(4\pi d_0)^2}
 \frac{(-1)^{1/4} e^{\frac{j (\rho -\rho')^2 k}{4(\tilde{\rho} -\tilde{\rho}')}} \sqrt{\pi} \left({\rm{Erfi}}\left[\frac{(-1)^{3/4} \sqrt{k} \left[(\rho -\rho') - (\tilde{\rho} -\tilde{\rho}') l_T\right]}{2 \sqrt{\tilde{\rho} -\tilde{\rho}'}}\right] -
    {\rm{Erfi}}\left[\frac{(-1)^{3/4} \sqrt{k} \left[(\rho -\rho') + (\tilde{\rho} -\tilde{\rho}') l_T\right]}{
    2 \sqrt{\tilde{\rho} -\tilde{\rho}'}}\right]\right)}{2 \sqrt{\tilde{\rho} -\tilde{\rho}'} \sqrt{k}}   
\end{equation}
\end{figure*}
\begin{figure*}[!ht]
\centering
  \begin{subfigure}[t]{.43\linewidth}
  \includegraphics[trim=0 0 0 0,clip,width=\linewidth]{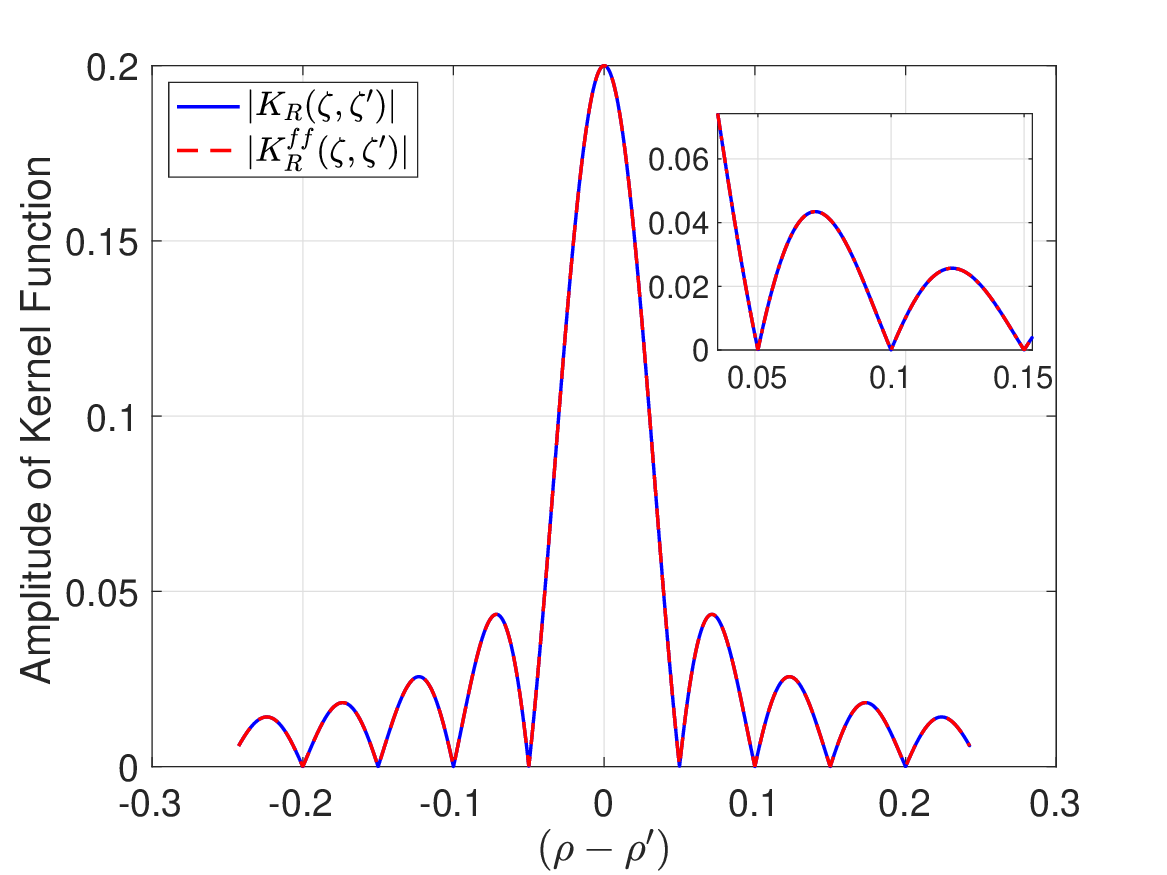}
\caption{$\theta_T=0,\theta_R=\pi,x_0=10\,m,y_0=0\,m,L_T=0.2\,m,L_R=5\,m,f=30\,$GHz.}
  \label{Kern_Comp_1a}
  \end{subfigure}\hfil
  \begin{subfigure}[t]{.43\linewidth}
    \includegraphics[trim=0 0 0 0,clip,width=\linewidth]{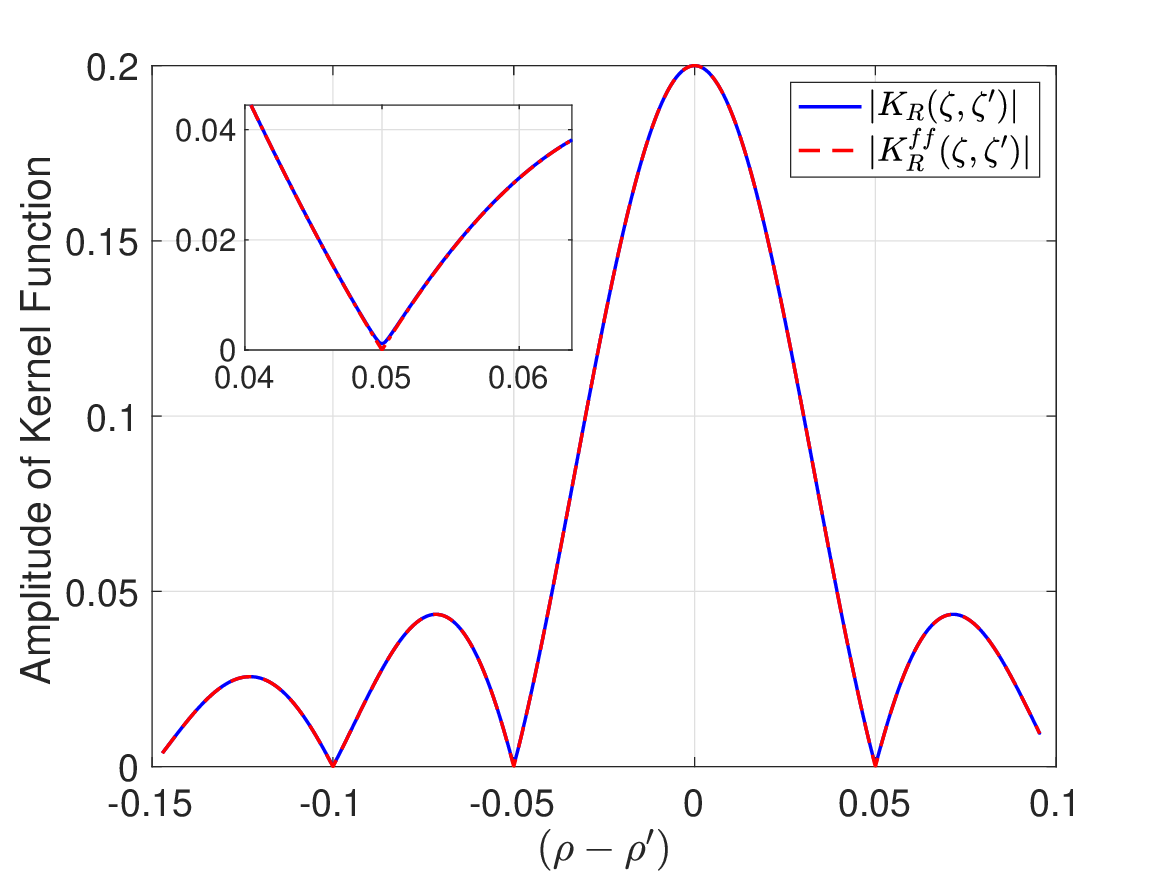}  \caption{$\theta_T=\pi/3,\theta_R=\pi,x_0=10\,m,y_0=0\,m,L_T=0.2\,m,L_R=5\,m,f=30\,$GHz.}
    \label{Kern_Comp_1b}
  \end{subfigure}\hfil
  \\
\begin{subfigure}[t]{.43\linewidth}
  \includegraphics[trim=0 0 0 0,clip,width=\linewidth]{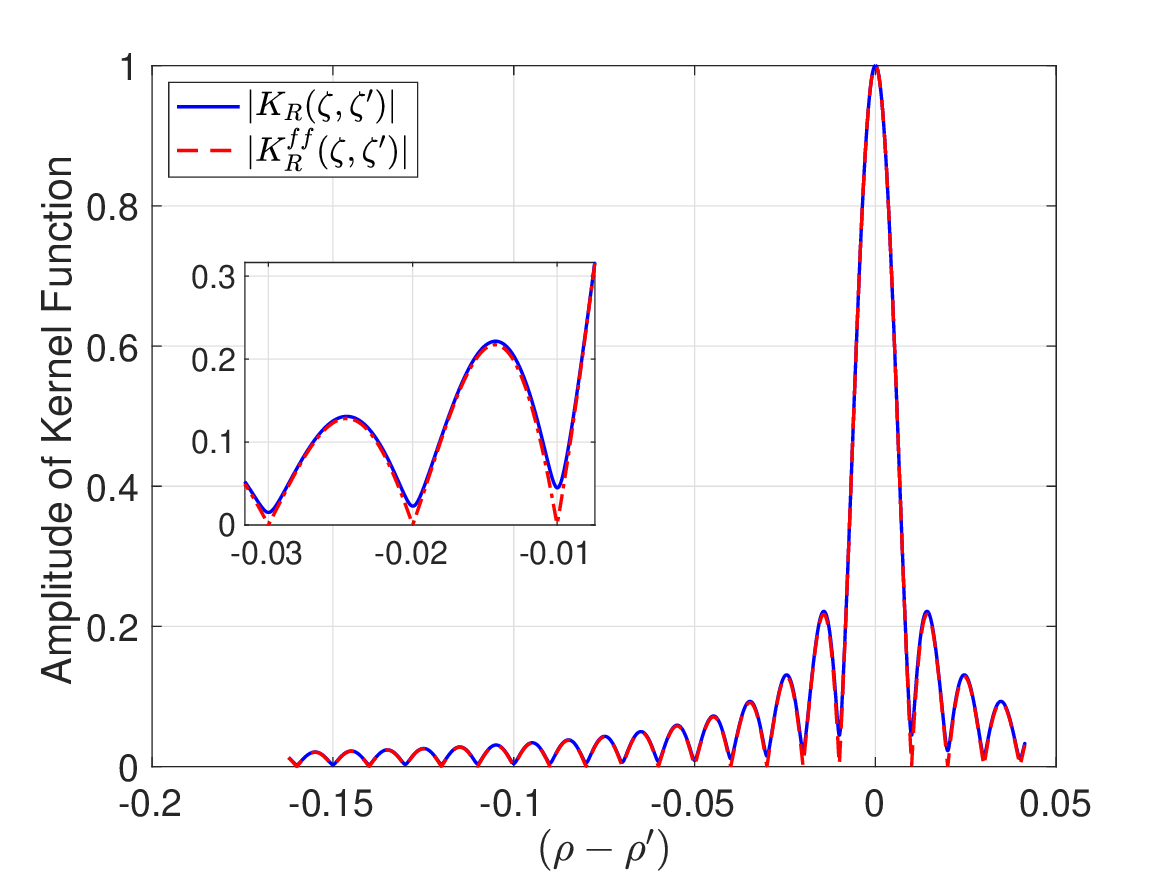}
   \caption{$\theta_T=\pi/3,\theta_R=-\pi/3,x_0=-5\,m,y_0=5\,m,L_T=1\,m,L_R=5\,m,f=30\,$GHz.}
  \label{Kern_Comp_1c}
  \end{subfigure}\hfil
  \begin{subfigure}[t]{.43\linewidth}
    \includegraphics[trim=0 0 0 0,clip,width=\linewidth]{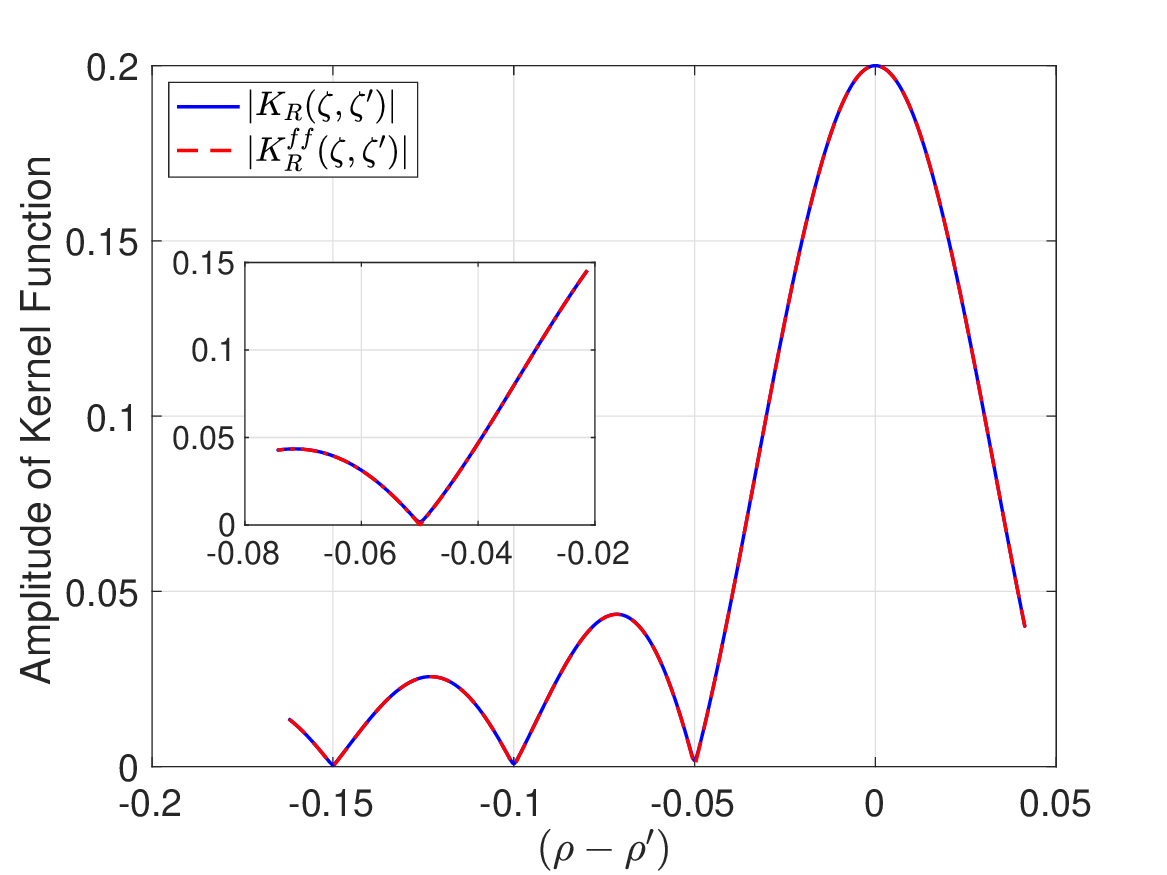}%
    \caption{$\theta_T=\pi/3,\theta_R=-\pi/3,x_0=-5\,m,y_0=5\,m,L_T=0.2\,m,L_R=5\,m,f=30\,$GHz.}
    \label{Kern_Comp_1d}
  \end{subfigure}\hfil
  
     \caption{Comparison of the amplitude of the kernel functions given by (\ref{kernel_3}) and (\ref{int_erf}) for four distinct cases. The inset is a zoomed picture to highlight the deviation of the two functions.}
  \label{Kernel_Comp_1}
\end{figure*}

In order to compute the number of DoF supported by the two arrays, we calculate the number of corresponding orthogonal focusing functions that fit within the length of the Rx array. This is accomplished by setting a reference point $\zeta'$ and finding the number and location of the points on the Rx array where the kernel function is zero. The analytical computation of the number of zeros on the Rx array for the function (\ref{int_erf}) is cumbersome. Fortunately, hereafter, it is shown that an alternative way to compute the number of zeros is to approximate the kernel function with the linear term of the phase, as if the Rx array were positioned in the far field of the Tx array. Under this assumption, the kernel function is given by
\begin{equation} \label{kernel_3}
K_R^{ff}(\zeta,\zeta') \approx  \frac{l_T}{(4\pi d_0)^2} {\rm{sinc}}\left(\frac{l_T}{\lambda}(\rho-\rho')\right),
\end{equation}
where ${\rm{sinc}}(x)=\sin(\pi x)/\pi x$, and the superscript $(ff)$ denotes the far field approximation. The usefulness of the ${\rm{sinc}\left(\frac{l_T}{\lambda}(\rho-\rho')\right)}$ function in \eqref{kernel_3} is that one may easily calculate the points $\zeta$, where it is zero, i.e., the integer multiples of $\frac{l_T}{\lambda}(\rho-\rho')$. Indeed, the amplitude of the kernel function in (\ref{int_erf}) oscillates in the same manner as the absolute value of the approximated kernel function in (\ref{kernel_3}) as a function of the difference $(\rho-\rho')$. Furthermore, the number of minima of the initial kernel and the approximated one is identical. This is clearly depicted in Fig. \ref{Kernel_Comp_1} where the two functions are plotted for different cases and $\zeta'=0$. The minima of the kernel functions occur at integer multiples of $\lambda/l_T$, where $l_T=L_T$ for the considered setups. It is noteworthy that when $(\rho-\rho') \gg (\tilde{\rho} -\tilde{\rho}')l_T$, as in Figs. \ref{Kern_Comp_1a}, \ref{Kern_Comp_1b} and \ref{Kern_Comp_1d} the approximation is highly accurate as the arguments of the ${\rm{Erfi}}$ functions in (\ref{int_erf}) are nearly identical. Consequently, the minima of the function $|K_R(\zeta,\zeta')|$ are congruent to those of the function $|K_R^{ff}(\zeta,\zeta')|$ and gradually approach zero. Interestingly, when $(\rho-\rho') > (\tilde{\rho} -\tilde{\rho}')l_T$, as in Fig. \ref{Kern_Comp_1c}, the number of minima remains the same, but their value are not necessarily zero for all of them. For the initial minima, i.e., for values of $\zeta$ close to the center of the Rx array, the deviation is greater, whereas for larger values of $\zeta$ it tends to zero. This behavior implies that some focusing functions are semi-orthogonal. However, since the effective length, $l_T$, of the Tx array plays a pivotal role on the ratio $[(\rho-\rho') /(\tilde{\rho} -\tilde{\rho}')l_T]$, the smaller the length, the smaller the deviation of the minima from the zero value.

It is worth noting that the proposed approach generalizes the work in \cite{Dardari21a} by  extending the phase profile with a non-linear term and allowing the Rx array: i) to rotate and ii) its center coordinate, $x_0$, to take negative values. Moreover, the Tx and Rx arrays radiate only in one of the two half-planes. 
To this end, we set the reference point at the center of the Rx array, i.e., $\zeta'=\zeta_c$, and solve the following equation for the points $\zeta_m$, which are measured with respect to the center $\zeta_c$ and fall within the array length, $-l_R/2\leq \zeta_m \leq l_R/2$:
\begin{equation}
\frac{l_T}{\lambda}(\rho_m-\rho_c) = m, \,\,\, m=\pm1,\pm2,\pm3,... 
\label{DoF_1}
\end{equation}
where $\rho_m$ is defined in \eqref{Taylor_2}, as follows:
\begin{equation}\label{rhom}
\begin{aligned}
    &\rho_m = \frac{\sin \theta_T - \gamma_m \cos \theta_T}{\sqrt{1+\gamma_m^2}} = \sin(\theta_T - a_m),\\
    &\gamma_m = \frac{y_0 + \zeta_c \cos \theta_R + \zeta_m \cos \theta_R - \eta_c \cos \theta_T}{x_0 - \zeta_c \sin \theta_R - \zeta_m \sin \theta_R + \eta_c \sin \theta_T} = \tan a_m.
\end{aligned}
\end{equation}
From the second identity in \eqref{rhom}, we obtain
\begin{equation} \label{zetam}
\begin{aligned}
&\zeta_m =\\
& \frac{\gamma_m (x_0-\zeta_c \sin \theta_R + \eta_c \sin \theta_T)-y_0-\zeta_c \cos \theta_R + \eta_c \cos \theta_T}{\gamma_m \sin \theta_R + \cos \theta_R}
\end{aligned}
\end{equation}
and by inserting (34) into the first identity in \eqref{rhom}, we obtain
\begin{equation}\label{gammam}
    \gamma_m = \tan a_m = \tan \left(\theta_T - \arcsin \left(\rho_c + m\frac{\lambda}{l_T}\right)\right).
\end{equation}
Substituting (\ref{gammam}) in (\ref{zetam}) and solving for $m$ at $\zeta_{m_+} = \frac{l_R}{2}$ and $\zeta_{m_-}=\frac{-l_R}{2}$, i.e., at the two extreme points of the Rx array, we obtain the two indices $m_+$ and $m_-$, as follows:
\begin{equation}\label{DoFplusminus}
\begin{aligned}
    & m_+ = \frac{l_T}{\lambda} [\sin (\theta_T - a_+)-\rho_c], \\
    & m_- = \frac{l_T}{\lambda} [\sin (\theta_T - a_-) -\rho_c],
\end{aligned}
\end{equation}
where the values of $a_+$ and $a_-$ are determined from the corresponding values of $\gamma_+$ and $\gamma_-$ that are obtained from (\ref{rhom}) by setting $\zeta_{m_+} = \frac{l_R}{2}$ and $\zeta_{m_-}=\frac{-l_R}{2}$, respectively. In addition, $\rho_c$ and $\gamma_c$ are calculated by setting $\zeta_m=0$ in (\ref{rhom}). 
Therefore, the number of DoF is given by
\begin{equation}\label{DoFs}
    m = |m_+ - m_-|+1.
\end{equation}
The constant 1 is added to ensure that the reference position is also taken into account, i.e., the mode that corresponds to $(\rho-\rho')=0$. In Fig. \ref{fig_All_ms}, a typical calculation of the number of DoF is illustrated as a function of $\theta_R$, based on (\ref{DoFs}).
\subsection{Insights on the Calculation of the Number of DoF}
The calculation of the number of DoF as outlined in the preceding section is explicitly dependent not only on the relative positions of the two arrays, as determined by the coordinates $(x_0, y_0)$, but also on their respective lengths $L_T, L_R$, and on the rotation angles $\theta_T$ and $\theta_R$. In this section, a concise analysis of the dependency of the number of DoF on all the aforementioned variables is presented. This analysis aims to elucidate the practical implications of specific constraints on the number of DoF.

\begin{figure}
\centering
  \includegraphics[width=0.9\linewidth]{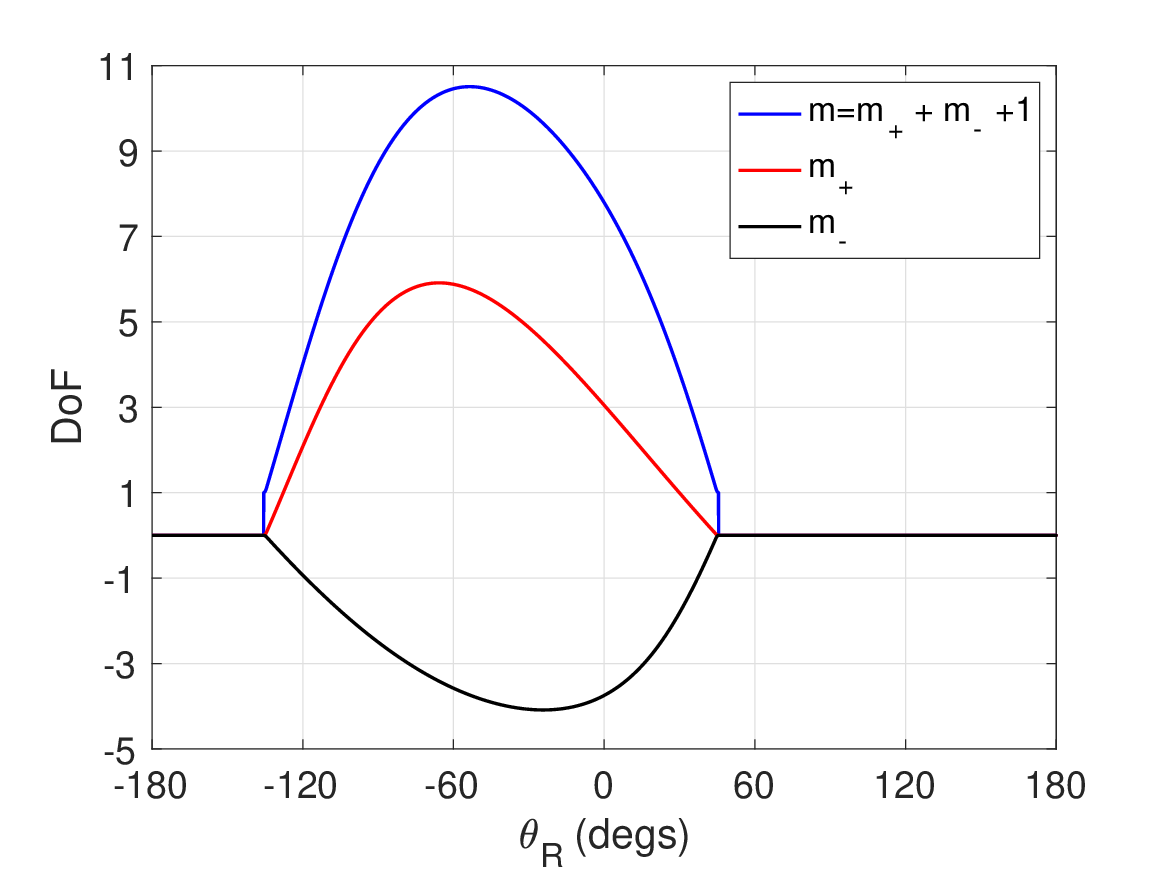}
  \caption{The DoF for $\theta_T=\pi/2$, $x_0 = -5$ m, $y_0 = 5$ m, $L_T = 0.2$ m, $L_R = 5$ m, $f = 30\,$GHz as given by (\ref{DoFs}).}
  \label{fig_All_ms}
\end{figure}

\begin{figure}
\centering
  \includegraphics[width=0.9\linewidth]{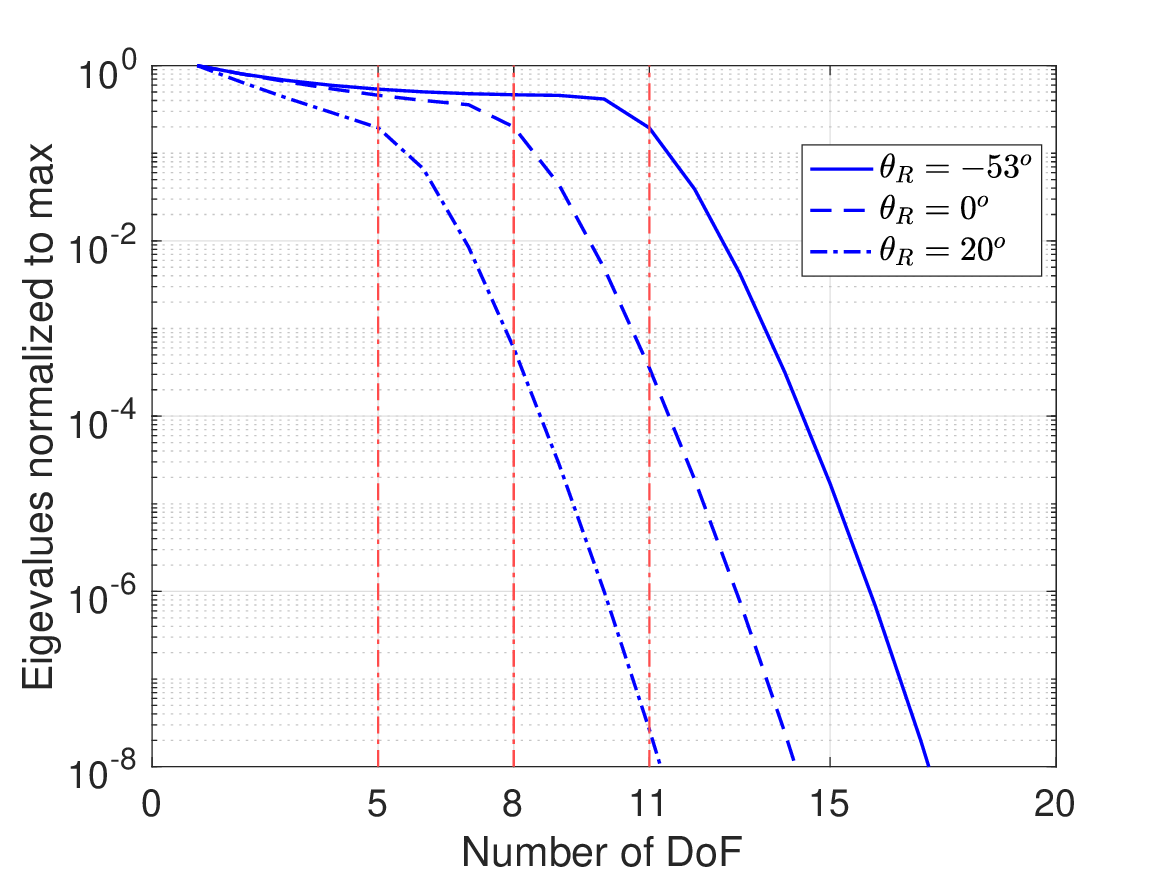}
  \caption{The DoF for $\theta_T=\pi/2$, $x_0 = -5$ m, $y_0 = 5$ m, $L_T = 0.2$ m, $L_R = 5$ m, $f = 30\,$GHz as calculated using the SVD of the LoS channel.}
  \label{DoF_SVD_1}
\end{figure}

The initial observation is that the variables $x_0$ and $y_0$ need to be selected to ensure $d_0=\sqrt{x_0^2+y_0^2}>1.2(L_T+L_R)$, for the approximation $r\approx r' \approx d_0$ for the amplitude in (\ref{kernel_1}) to be valid. This constraint does not impose a significant restriction on the selection of the pair of $x_0,y_0$, as for a small Tx array and a large Rx array, the lower limit for the distance $d_0$ is relatively small. Furthermore, this constraint contributes to the desired value of the ratio $[(\rho-\rho’) /(\tilde{\rho} -\tilde{\rho}’)l_T]$, i.e., the greater the distance $d_0$, the greater the ratio becomes. The second observation is that the larger the values of $L_T$ and $L_R$, the larger the number of DoF, as expected and also shown in Fig. \ref{Kern_Comp_1c} and \ref{Kern_Comp_1d}. However, the value of $L_T$ also determines the effective length $l_T$ that is desirable to keep it at relatively low values. If $L_T \ll d_0$, the ratio $[(\rho - \rho’) / (\tilde{\rho} - \tilde{\rho}’)l_T]$ is kept large and achieves the orthogonality of the focusing functions.
The third observation is that the rotation angles, $\theta_T$ and $\theta_R$, not only determine the visibility of the two arrays but also alter the projected lengths of the two linear arrays onto the perpendicular line connecting the centers of the two arrays. This has a direct consequence on the number of DoF as also observed by comparing Figs. \ref{Kern_Comp_1a} and \ref{Kern_Comp_1b}. The partial visibility of the Tx or Rx arrays directly affects the number of DoF in the specific geometric setup. In fact, the partial visibility of the Tx array implies a reduced length compared to $L_T$, which in turn changes the position of the zeros of the kernel on the Rx array. Furthermore, the number of DoF increases with $l_T$ according to (\ref{DoF_1}). The same applies to the reduced effective length of the Rx array, which affects the range of the $\zeta$ variable and, consequently, the number of DoF.
Using the LoS channel model and performing a singular value decomposition (SVD) to the channel matrix, one can obtain the DoF for a specific geometric setup of the two arrays. The channel matrix is a non-square matrix that contains the values of the scalar free space Green’s function that models the response at a point, $\zeta_m$, on the Rx array from a wave transmitted from a point $\eta_n$ on the Tx array. Specifically, using the retarded solution to the wave equation, i.e., the outgoing wave, one may write $G(\zeta_m,\eta_n)=\frac{1}{4\pi}\frac{\exp{(-jkr_{mn}})}{r_{mn}}$, where $r_{mn}$ is given by (\ref{dist}) for $\zeta=\zeta_m$ and $\eta=\eta_n$. Therefore, the distance from a Tx point to an Rx point is calculated without any approximation. Fig. \ref{DoF_SVD_1} presents the eigenvalues normalized to the maximum eigenvalue versus their index, for the set of parameters used in Fig. \ref{fig_All_ms}, and for three values of the angle $\theta_R$. It is evident that there is a threshold for the number of DoF after which the eigenvalues decrease rapidly. The precise number of DoF that corresponds to the knee of the curves depends on the power coupling strength of the modes. The sum rule on the eigenvalues provides the total power strength of the eigenvalues and by setting a percentage of the sum rule to be guaranteed, one can determine the number of orthogonal modes to retain \cite{Miller19}. Orthogonality is directly related to the zeros of the kernel function, through the inner products in the Hilbert spaces. The eigenvalues of the modes after the knee indicate a weak strength and the inability to exploit additional DoF. Consequently, the knee corresponds to the effective “diffraction” limit, or the so-called effective number of DoF, and its value almost coincides with the value displayed in Fig. \ref{fig_All_ms} for the corresponding values of $\theta_R$ and for a threshold level of 99\% of the sum rule. The 11th eigenvalue for $\theta_R=53^o$ is relatively weak, $|s_{11}|^2=0.194$, compared to the 10th, $|s_{10}|^2=0.416$, where $|s_j|$ denotes the magnitude of the $j$-th singular value. The percentage of cumulative strength of the first ten eigenvalues in approximately 96\%.

\subsection{A Geometric Setup of Interest}
\begin{figure*}[!ht]
  \begin{subfigure}[t]{.33\linewidth}
  \includegraphics[trim=0 0 0 0,clip,width=\linewidth]{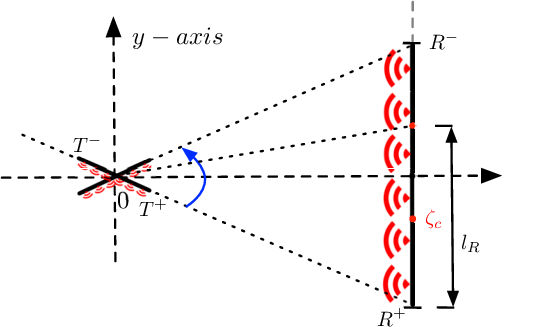}
   \caption{The first branch of (\ref{Theta_T})}
  \label{Branch_1}
  \end{subfigure}\hfil
  \begin{subfigure}[t]{.33\linewidth}
    \includegraphics[trim=0 0 0 0,clip,width=\linewidth]{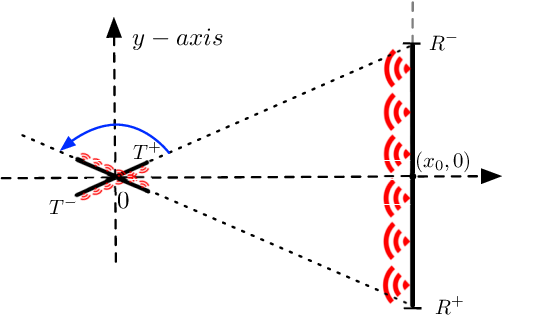}%
    \caption{The second branch of (\ref{Theta_T})}
    \label{Branch_2}
  \end{subfigure}\hfil
  \begin{subfigure}[t]{.33\linewidth}
    \includegraphics[trim=0 0 0 0,clip,width=\linewidth]{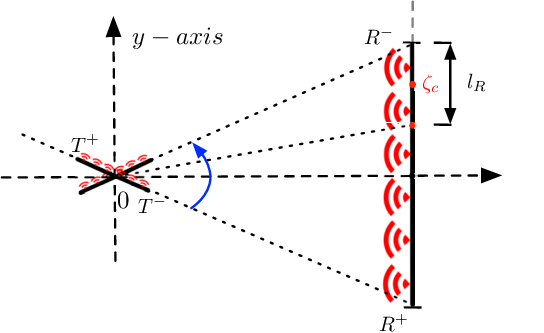}%
    \caption{The third branch of (\ref{Theta_T})}
   \label{Branch_3}
  \end{subfigure}%
   \caption{The range of $\theta_T$ that results in partial and full Rx visibility according to (\ref{Theta_T})}
  \label{ThetaT_Branches}
\end{figure*}

In this subsection, a practical geometric setup is considered where a small array is located in front of a large array, which resembles the placement of user equipment in front of a radio unit. This case study is of interest since, for a range of values of the variables $x_0$ and $\theta_T$, it provides a paraxial setup, i.e., the midpoints of the two arrays are perfectly aligned along the same line and thus $y_0 = 0$, whereas for other values of $x_0$ and $\theta_T$ the center of the Rx is shifted, leading to a non-paraxial case. This is due to the partial visibility of the Rx array and thus the need to calculate the DoF with respect to the new center. This setup will also be utilized in the next section to study the statistical behavior of the DoF.
Moreover, assume that $x_0 \geq 1.2(L_T+L_R)$, that is, Rx is in the right half-plane, and $\theta_R=\pi$. The coordinate $x_0$ and the rotation angle of the TX surface determine the visibility condition, and consequently the number of DoF between the two surfaces. Therefore, there are four possible cases: i) the Rx array is not visible from the Tx and the DoF is 0, ii) the Rx array is partially visible from the Tx array and the endpoint $R^+$ is visible, iii) the Rx array is fully visible from the Tx, and iv) the Rx array is partially visible from the Tx and the endpoint $R^-$ is visible. The first case occurs for
$-\pi < \theta_T \leq a_+^{max} -\frac{\pi}{2}$ and for $a_-^{max} + \frac{\pi}{2} \leq \theta_T \leq \pi$, where $a_+^{max}=\arctan \gamma_+ =\arctan (-L_R/2 x_0)$ and $a_-^{max} = \arctan \gamma_- = \arctan (L_R/2 x_0) = -a_+^{max}$. Then, 
\begin{equation}\label{Theta_T}
    \mbox{if} \left\{ \begin{array}{rcl} a_+^{max} - \frac{\pi}{2} < \theta_T < a_-^{max} - \frac{\pi}{2} &  \mbox{partial vis., $R^+$ visible} \\
    a_-^{max} - \frac{\pi}{2} \leq \theta_T \leq a_+^{max} + \frac{\pi}{2} & \mbox{full visibility} \\
    a_+^{max} + \frac{\pi}{2} < \theta_T < a_-^{max} + \frac{\pi}{2} & \mbox{partial vis., $R^-$ visible}
    \end{array}\right.
\end{equation}
The three branches are depicted in Fig. \ref{ThetaT_Branches}. For the first and third branches, we have depicted a random selection of $\theta_T$ along with the new length $l_R$ and the new center $\zeta_c$ of the Rx surface. For the middle branch of full visibility, the setup implies that $a_0 = 0$ and consequently $\rho_c=\sin \theta_T$. The case $\zeta_m = L_R/2$, yields $a_+ = a_+^{max}$, and the case $\zeta_m = -L_R/2$ yields $a_- = a_-^{max} = -a_+^{max}$. Note that for this setup, $a_- > 0$ and $a_+ < 0$. Substituting into \eqref{DoFplusminus}, it can be rewritten as  
\begin{equation} \label{mplus-paraxial}
   m_+ = \frac{L_T}{\lambda} [\sin \left(\theta_T - a_+ \right)
   - \sin \theta_T],
\end{equation}
\begin{equation}
   m_- = \frac{L_T}{\lambda} [\sin \left(\theta_T - a_- \right)
   - \sin \theta_T].
\end{equation}
Then, the number of DoF under the full visibility scenario is 
\begin{equation}\label{DoF_case}
\begin{aligned}
m &= 1+\frac{L_T}{\lambda} [\sin \left(\theta_T - a_+ \right) - \sin \left(\theta_T - a_- \right)]\\
&\mathop=\limits^{a_+=-a_-} 1+\frac{2 L_T}{\lambda} \cos \theta_T
 \sin a_-.    
\end{aligned}
\end{equation}
The absolute value in (\ref{DoFs}) is omitted, since $\sin \left(\theta_T - a_+ \right)>0$ and $\sin \left(\theta_T - a_- \right)<0$. 
For a very large receiving surface, i.e., for $L_R \rightarrow\infty$ and in the paraxial case, i.e., $\theta_T=0$, $y_0=0$, the limit is $\lim_{L_R\to\infty}m \approx 2 L_T/\lambda$.
For the first branch of (\ref{Theta_T}), the Rx array is crossed and the $R^+$ endpoint is always visible. For the third branch, on the other hand, the $R^-$ endpoint is always visible. Then, (\ref{intersectionpoint}) can be used for the calculation of the intersection point, which depends on the coordinate $x_0$ and the angle $\theta_T$. This allows for the calculation of the new length of Rx using (\ref{Rxlength}). To this end, the new angles should be calculated as follows:
\begin{equation} \label{newalpha}
\begin{aligned}
     a_0 & = \arctan \left(\frac{y_0+\zeta_c \cos \theta_R}{x_0 -\zeta_c \sin \theta_R}\right) = \arctan \left(\frac{-\zeta_c}{x_0}\right), \\
     a_+ & = \arctan \left(\frac{y_0+(\zeta_c + \frac{l_R}{2}) \cos \theta_R}{x_0 -(\zeta_c + \frac{l_R}{2}) \sin \theta_R}\right) \\
     &= \arctan \left(\frac{-\zeta_c-\frac{l_R}{2}}{x_0}\right), \\
\end{aligned}
\end{equation}
\begin{equation*}
\begin{aligned}
a_- & = \arctan \left(\frac{y_0+(\zeta_c - \frac{l_R}{2}) \cos \theta_R}{x_0 -(\zeta_c - \frac{l_R}{2}) \sin \theta_R}\right) \\
&= \arctan \left(\frac{-\zeta_c+\frac{l_R}{2}}{x_0}\right),
\end{aligned}
\end{equation*}
where $\zeta_c$ is calculated using (\ref{Rxcentre}). Also, $\rho_c=\sin (\theta_T - a_0)$ and the calculation of the number of DoF is obtained from (\ref{DoFplusminus}) for $l_T=L_T$.

\section{Statistical Analysis of the DoF}
In this section, a stochastic geometry framework is used to study the statistical behavior of the DoF in a mmWave network under a simple but practical scenario. Although in the general case there are four random variables, that is, $x_0, y_0, \theta_T, \theta_R$, that determine the spatial characteristics of the network setting considered, only 2 parameters, $x_0, \theta_T$ are random variables for the practically relevant scenario considered in Subsection IV.A. Owing to the statistical dependence across four random variables, we will focus on this case study to maintain tractability. 

\subsection{Stochastic Geometry Framework}
We consider a wireless network, where the spatial location of the center $(x_0,y_0)$ of the receiving array is modeled as a uniform binomial point process (BPP) $\Phi$, in a finite region $\mathcal{A} \subset \mathbb{R}^2$. Without loss of generality, it is assumed that $\mathcal{A}= \mathbf{b}(\mathbf{o},R)$, where $\mathbf{b}(\mathbf{o},R)$ denotes a ball of radius $R$ centered at the origin $\mathbf{o}$. In addition, we assume that the Tx array is located at $\mathbf{o}$ and that $y_0=0$. The Tx and Rx arrays may be fully visible, partially visible, or not visible at all. Under these assumption, the number of DoF in \eqref{DoFs} simplifies as follows:
\begin{equation}\label{S1}
m = 1+\frac{L_T}{\lambda}|\sin (\theta_T - a_+ ) - \sin(\theta_T - a_- )|= 1+\frac{L_T}{\lambda} |\rho_{+}-\rho_{-}|.
\end{equation}

In the following, the statistical analysis of the DoF is presented for two case studies: i) $x_0$ is a random variable and ii) analysis conditioned on $x_0$. Statistical analysis is performed as a function of the visibility between the two arrays. 

\subsection{The Coordinate $x_0$ is a Random Variable}
In this subsection, the number of DoF is evaluated under the full and partial visibility conditions for the Rx array.

\subsubsection{Partial Visibility}
For this scenario, the number of DoF is given by \eqref{S1}, where $a_+$ and $a_-$ are given in \eqref{newalpha}. Depending on the angle $\theta_T$, either the point $R^{+}$ or the point $R^{-}$ of the Rx array may be visible. These case studies are formulated in \eqref{Theta_T} with respect to the angle $\theta_T$. Both cases are analyzed. 
The PDF of the number of DoF is first derived under the assumption that $R^{+}$ is visible. Interestingly, for this case, $\rho_{+}$ and $\rho_{-}$ are statistically independent, as shown in the following lemma.

\emph{Lemma 1:} \emph{ Assuming partial visibility of the Rx array and the point $R^+$ being visible, the function $\rho_{+}$ which depends on the random variables $\theta_T$ and $x_0$, takes a constant value equal to $\rho_{+}=-1$, and the PDF $f_{\rho_{-}}(\rho)$ of the random variable $\rho_{-}$ is given by }
\begin{equation}\label{Lemma1}
f_{\rho_{-}}(\rho) = \int_{0}^{[x_{max}(\rho)]^{-}} \frac{1}{2\sqrt{1-\rho^2}}  \frac{1}{\arctan(\frac{L_R}{2 x_0})} f_{x_0}(x_0) {\rm{d}} x_0, 
\end{equation}
\emph{where $\rho_{-} \in [-1,1]$, $f_{x_0}(x_0) = \frac{4 \sqrt{R^2-x_0^2} }{\pi R^2}$, $x_{max}(\rho)=\frac{L_R}{2 \tan \big(\frac{\arcsin \rho + \frac{\pi}{2}}{2}\big)}$ and $[x]^{-} = \min\{x,R\}$.}

\textit{Proof.} See Appendix A. \QEDA  

Having obtained $\rho_{+}$ and $\rho_{-}$, the PDF of the number of DoF is derived in the following theorem.

\emph{Theorem 1:} \emph{Under partial visibility of the Rx array when $R^{+}$ is visible, the PDF $f_{m|R^{+}}(m)$ of the number of DoF, is 
\begin{equation}\label{Theorem1}
f_{m|R^{+}}(m) =  \frac{1}{C_{L_T, \lambda}} f_{\rho_{-}}\Big(\frac{m}{C_{L_T, \lambda}}-1\Big) 
\end{equation}
for $m \in [0, 2 C_{L_T, \lambda}]$, where $C_{L_T, \lambda}=\frac{ L_T}{\lambda}$.}

\textit{Proof.} The proof follows directly from Lemma 1 through the change of variables $\rho_{-}=\frac{m}{C_{L_T, \lambda}}+\rho_{+}$ to satisfy \eqref{DoFs} and due to the independence of $\rho_{+}$ from $\rho_{-}$. \QEDA 

In the following, the PDF of the number of DoF is derived under the assumption that the point $R^-$ is visible. 

\emph{Lemma 2:} \emph{ Assuming partial visibility of the Rx array and the point $R^-$ being visible, the function $\rho_{-}$, which depends on the random variables $\theta_T$ and $x_0$, takes a constant value equal to $\rho_{-} = 1$, and the PDF $f_{\rho_{+}}(\rho)$ of the random variable $\rho_{+}$ is given by }
\begin{equation}
\begin{split}
&f_{\rho_{+}}(\rho) \\
&= \int_{0}^{[x_{max}(f(\rho))]^{-}} \frac{1}{2\sqrt{1-\rho^2}}  \frac{1}{\arctan(\frac{L_R}{2 x_0})} f_{x_0}(x_0) {\rm{d}} x_0, 
\end{split}
\end{equation}
\emph{where $\rho_{+} \in [-1,1]$ and $f(\rho)=\sin (\arccos \rho -\frac{\pi}{2}) = -\rho$.}

\textit{Proof.} By recalling \eqref{newalpha}, $\alpha_{+}$ can be simplified to $\alpha_{+}=\arctan\big(-\frac{L_R}{2 x_0}\big)$ and $\alpha_{-} = \arctan (-\cot \theta_T)$. Next, the proof follows similar conceptual steps as in Appendix A, and it is hence omitted here for brevity. \QEDA

Having obtained $\rho_{+}$ and $\rho_{-}$, the PDF of the number of DoF is derived in the following theorem.

\emph{Theorem 2:} \emph{Under partial visibility of the Rx array with the point  $R^{-}$ being visible, the PDF $f_{m|R^{-}}(m)$ of the number of DoF, is 
\begin{equation}\label{Theorem2}
\begin{split}
f_{m|R^{-}}(m) &=  \frac{1}{C_{L_T, \lambda}} f_{\rho_{-}}\Big(1-\frac{m}{C_{L_T, \lambda} }\Big)\\
&= \frac{1}{C_{L_T, \lambda}} f_{\rho_{+}}\Big(\frac{m}{C_{L_T, \lambda} }-1\Big),
\end{split}
\end{equation}
for $m \in [0, 2 C_{L_T, \lambda}]$.}

\textit{Proof.} The proof follows directly from Lemma 2 through the change of variables $\rho_{+}=\rho_{-}-\frac{m}{C_{L_T, \lambda} }$ to satisfy \eqref{DoFs} and due to the independence of $\rho_{+}$ from $\rho_{-}$. \QEDA

\subsubsection{Full Visibility of the Receiving Array}
Under the full visibility scenario, the number of DoF is given by \eqref{DoF_case}, where $a_- = \arctan(\gamma_-)$. Conditioned on $x_0$, the conditional PDF $f_{m|x_0}(m)$ is first derived.

\emph{Lemma 3:} \emph{Assuming that the Rx array is fully visible ($FV$), the conditional PDF $f_{m|x_0}(m)$ of the number of DoF, is}
\begin{equation}\label{Lemma3}
\begin{split}
&f_{m|x_0}(m)= \\
& \frac{1}{  C_{L_T, \lambda}(x_0)}  \frac{1}{\sqrt{1-\big(\frac{m}{C_{L_T, \lambda}(x_0)}\big)^2}} \frac{1}{\pi-2\arctan\big(\frac{L_R}{2 x_0}\big)},
\end{split}
\end{equation}
\emph{ for $m \in \Big[ \frac{2 C_{L_T, \lambda}  L_R^2}{L_R^2+ 4 x_0^2}, 2C_{L_T, \lambda}(x_0) \Big]$ and $C_{L_T, \lambda}(x_0) =  C_{L_T, \lambda} \sin \big(\arctan\big(\frac{L_R}{2 x_0}\big)\big)$.}

\textit{Proof.} Conditioned on $x_0$ and recalling \eqref{DoF_case}, it is observed that $a_-$ is no longer a random variable. Next, the proof follows by applying transformations upon the single random variable $\theta_T$. \QEDA 

Then, the PDF $f_{m}(m)$ of the number of DoF under the full visibility assumption is derived with the aid of Lemma 3 as follows. 

\emph{Theorem 3:} \emph{Under full visibility of the Rx array, the PDF $f_{m}(m)$ of the number of DoF, is given by }
\begin{equation}\label{Theorem3}
\begin{split}
&f_{m|FV}(m)= \int_{[\omega(m)]^{-}}^{[\psi(m)]^{-}} f_{m|x_0}(m) f_{x_0}(x_0) {\rm{d}} x_0,
\end{split}
\end{equation}
\emph{for $m \in [0,2 C_{L_T, \lambda}]$, where $\omega(m)=\sqrt{\frac{2 C_{L_T, \lambda} L_R^2-m L_R^2}{4 m}}$ and $\psi(m) = \frac{L_R}{2 \tan\Big(\sin^{-1}\Big(\frac{m}{2 C_{L_T, \lambda}}\Big)\Big)}$}

\textit{Proof.} By deconditioning the PDF $f_{m|x_0}(m)$ over $x_0$, the PDF $f_{m|FV}(m)$ is given by
\begin{equation}
\begin{split}
&f_{m|FV}(m)= \int_{0}^{R} f_{m|x_0}(m) f_{x_0}(x_0) {\rm{d}} x_0,
\end{split}
\end{equation}
where $m \in \Big[ \frac{2 C_{L_T, \lambda} L_R^2}{L_R^2 + 4 x_0^2}, 2 C_{L_T, \lambda} \sin(\tan^{-1}(\frac{L_R}{2 x_0}))\Big]$. Finally, we solve $\frac{2 C_{L_T, \lambda} L_R^2}{L_R^2 + 4 x_0^2}=m$ and $2 C_{L_T, \lambda} \sin(\tan^{-1}(\frac{L_R}{2 x_0}))=m$ for $x_0$, making the ranges of $m$ independent of $x_0$. This concludes the proof. \QEDA

\subsection{Analysis Conditioned on $x_0$}
In this subsection, the statistical analysis of the number of DoF is conducted by conditioning on the coordinate $x_0$ of the center of the Rx array.
Conditioned on $x_0$, there is a probability that the Rx array is visible, which is formally defined as follows.

\emph{Definition 1 (Probability of Visibility (PoV)):} \emph{The PoV $\mathcal{V}$ is defined as the probability that the Rx array is either partially or fully visible from the Tx array. The PoV is given by $\mathcal{V} = \frac{1}{2} + \frac{\arctan\big(\frac{L_R}{2 x_0}\big)}{\pi}$.}

\textit{Proof.} The PoV $\mathcal{V} = \frac{\mathbb{P}[\mathrm{Partial \, or \, Full \, Visibility}]}{2 \pi}$. Recalling \eqref{Theta_T}, the probability $\mathbb{P}[\mathrm{Partial \, or \, Full \, Visibility}] = a_-^{max} + \frac{\pi}{2} - a_+^{max} + \frac{\pi}{2} = \pi + 2\arctan(L_R/2 x_0)$. By substituting $\mathbb{P}[\mathrm{Partial \, or \, Full \, Visibility}]$ in $\mathcal{V}$ the proof is complete. \QEDA

Next, the conditional PDFs of the number of DoFs with partial or full visibility are derived as follows. 

\subsubsection{Partial Visibility of the Rx Array}

Conditioned on $x_0$ and as the Tx array rotates of an angle $\theta_T$, either the $R^{+}$ or the $R^{-}$ point of the Rx array may be visible, each with a given probability. Assuming partial visibility, the conditional PDF of the number of DoF is given in the following lemma.

\emph{Lemma 4:} \emph{Conditioned on $x_0$ and $R^+$ or $R^-$ being visible, the conditional PDF $f_{m|x_0,R^{+}}(m)$ and $f_{m|x_0,R^{-}}(m)$ of the number of DoF, $m$, are respectively, given by}
\begin{equation}\label{Lemma4}
\begin{split}
&f_{m|x_0,R^{+}}(m) = f_{m|R^{+}}(m)\Big|_{f_{x_0}(x_0)=\delta(x_0)} ,\\
&f_{m|x_0,R^{-}}(m) = f_{m|R^{-}}(m)\Big|_{f_{x_0}(x_0)=\delta(x_0)},
\end{split}
\end{equation}
\emph{for $m \in \Big[0, 2 C_{L_T, \lambda} \frac{L_R^2}{L_R^2+ 4 x_0^2}\Big]$.}
\textit{Proof.} Assume that $R^+$ is visible. Conditioned on $x_0$, one can set $f_{x_0}(x_0)=\delta(x_0)$ in \eqref{Lemma1}, which provides the PDF $f_{\rho_{-}}(\rho)$. The limits of the range for $m$ are obtained with the aid of \eqref{AppAfA_} shown in Appendix A. Therefore, $f_{m|x_0,R^{+}}(m) = f_{m|R^{+}}(m)\Big|_{f_{x_0}(x_0)=\delta(x_0)}$ and similarly $f_{m|x_0,R^{-}}(m) = f_{m|R^{-}}(m)\Big|_{f_{x_0}(x_0)=\delta(x_0)}$. Finally, $f_{m|x_0,R^{+}}(m)$ and $f_{m|x_0,R^{i}}(m)$ have the same statistical behavior. \QEDA 
\subsubsection{Full Visibility of the Rx Array}
Conditioned on $x_0$, and with full visibility of the Rx array, the conditional PDF $f_{m|x_0,FV}(m)$ of the number of DoF is given by Lemma 3 as $f_{m|x_0,FV}(m) = f_{m|x_0}(m)$. 
Conditioned on $x_0$, the PDF of the number of DoF is given by the following theorem.
\emph{Theorem 4:} \emph{ Conditioned on $x_0$, the conditional PDF $f_{m|x_0}(m)$ of the number of DoF, $m$, is given by \eqref{conditional}, shown at the bottom of the next page,}
\emph{where $\mathcal{V}_{R^+} = \mathcal{V}_{R^-} = \frac{\arctan(L_R/2 x_0)}{\pi}$ denotes the probability that $R^{+/-}$ is visible and $\mathcal{V}_{FV} = \frac{\pi - 2\arctan(L_R/2 x_0)}{2 \pi}$ denotes the probability that full visibility is achievable.}
\begin{figure*}[!hb]
\hrulefill
\begin{equation}\label{conditional}
f_{m|x_0}(m)  = \left\{ 
\begin{array}{ll}
 \frac{\mathcal{V}_{R^+}}{\mathcal{V}} f_{m|x_0,R^{+}}(m) + \frac{\mathcal{V}_{R^-}}{\mathcal{V}} f_{m|x_0,R^{-}}(m),   & 0 \leq m \leq  \frac{2 C_{L_T, \lambda}L_R^2}{L_R^2+ 4 x_0^2} \\
   \frac{\mathcal{V}_{FV}}{\mathcal{V}} f_{m|x_0,FV}(m) , &   \frac{2 C_{L_T, \lambda}L_R^2}{L_R^2+ 4 x_0^2} < m < 2C_{L_T, \lambda}(x_0),
\end{array} 
\right.
\end{equation}
\end{figure*}
\textit{Proof.} The proof follows directly from Lemma 3 and Lemma 4 after calculating the probability of establishing partial or full visibility, conditioned on the visibility event. \QEDA

\emph{Remark 2:} \emph{Theorem 4 captures and quantifies the maximum number of DoF that can be supported under partial and full visibility, conditioned on the coordinate $x_0$. This becomes a key limiting factor in the achievable performance of mmWave networks if the user association policy is based on the maximum achievable number of DoF.}

\section{Numerical Results and Discussion}

\subsection{Deterministic Analysis}
\begin{figure*}[!ht]
  \begin{subfigure}[t]{.33\linewidth}
  \includegraphics[trim=0 0 0 0,clip,width=\linewidth]{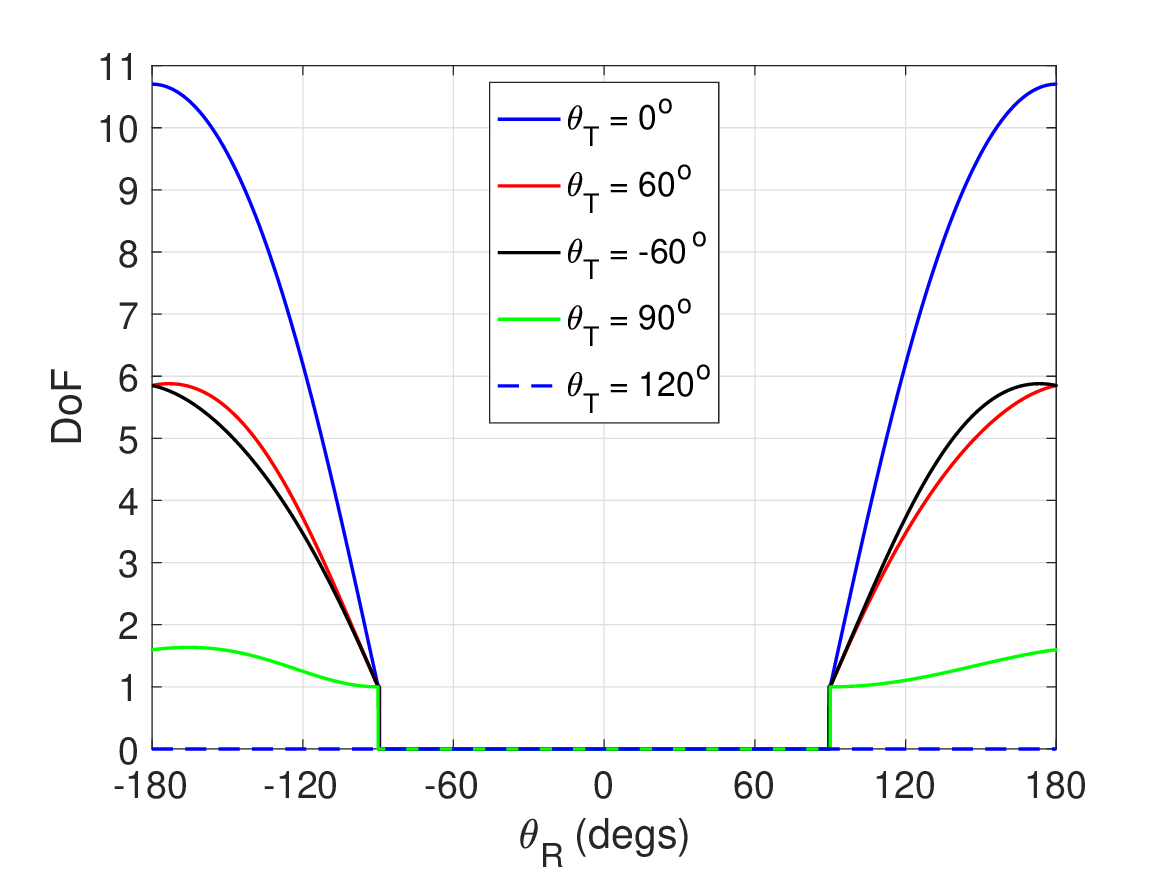}
   \caption{The DoF for $x_0 = 10m$, $y_0 = 0m$}
         \label{fig_NumRes_1a}
  \end{subfigure}\hfil
  \begin{subfigure}[t]{.33\linewidth}
    \includegraphics[trim=0 0 0 0,clip,width=\linewidth]{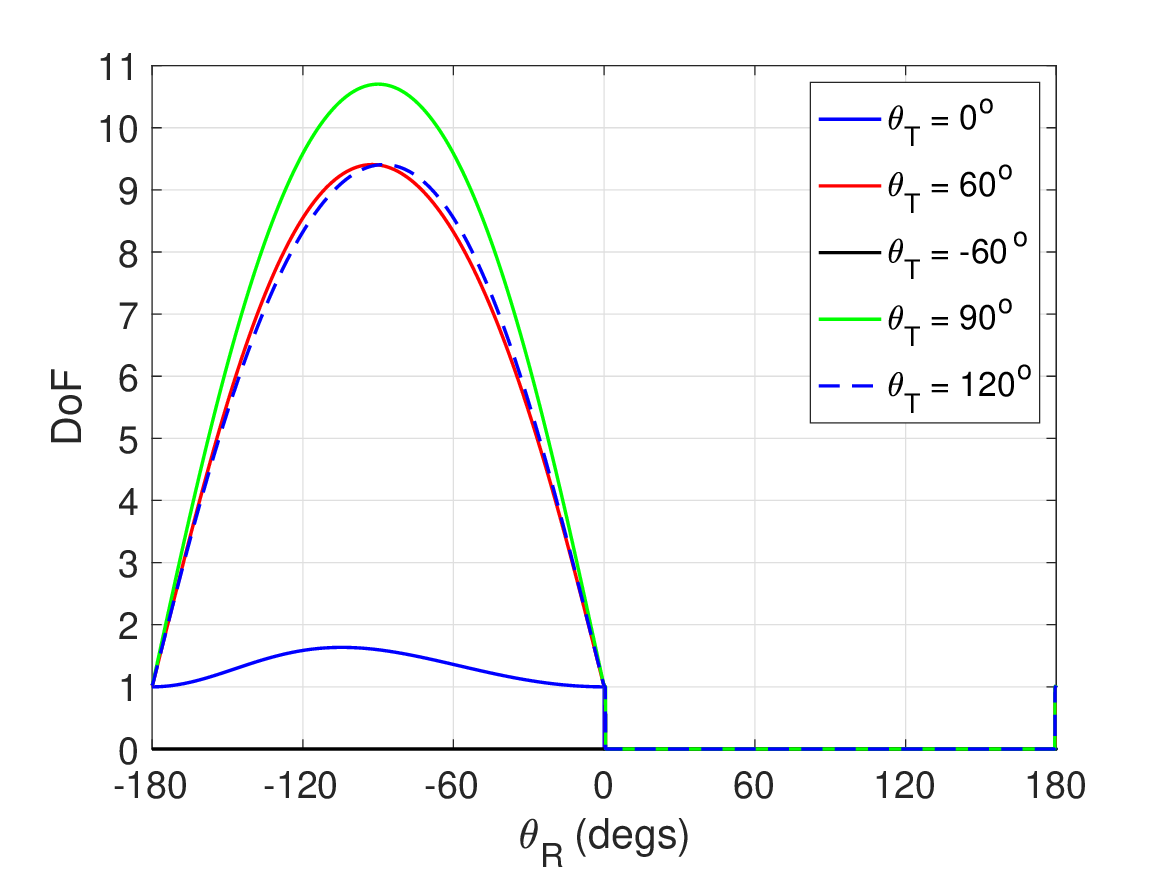}%
    \caption{The DoF for $x_0 = 0m$, $y_0 = 10m$}
         \label{fig_NumRes_1b}
  \end{subfigure}\hfil
  \begin{subfigure}[t]{.33\linewidth}
    \includegraphics[trim=0 0 0 0,clip,width=\linewidth]{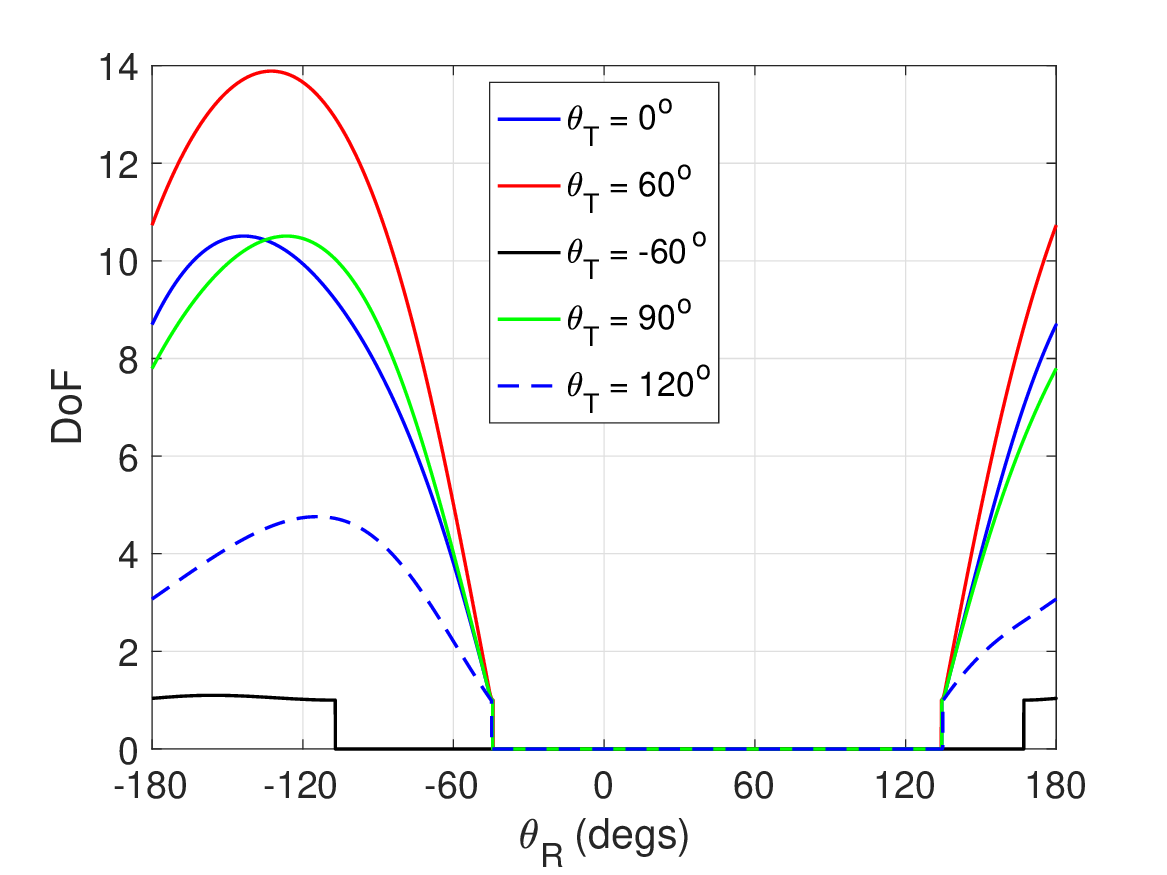}%
    \caption{The DoF for $x_0 = 5m$, $y_0 = 5m$}
         \label{fig_NumRes_1c}
  \end{subfigure}%
   \caption{The DoF for $L_T = 0.2m$, $L_R = 5m$, $f = 30\,$GHz.}
   \label{fig_NumRes_1}
\end{figure*}

Because there are seven variables involved in the calculation of the number of DoF, namely $\theta_T, \theta_R, x_0, y_0, L_T, L_R, f$, the number of possible combinations is extremely large. Thus, in this subsection, representative cases are examined and numerical results are given. In all results presented, the frequency is set to $f=30\,$GHz, and the length of the transmitting surface is set to $L_T = 0.2$ meters. In Fig. \ref{fig_NumRes_1}, the number of DoF for three different Rx locations is given assuming that a large receiving array of length $L_R = 5$ meters is used. The number of DoF is given versus $\theta_R \in (-\pi/2, \pi/2]$, for various values of $\theta_T$. The line that connects the centers of the two arrays is the $x$-axis in the first plot, the $y$-axis in the second plot, and it forms an angle of $\pi/4$ with the $x$-axis in the third plot. All the depicted values match the number of zeros of the kernel function as shown in Fig. \ref{Kernel_Comp_1}. The first observation is that a different maximum number of DoF is obtained for different values of $\theta_T$. 
Notably, the maximum value of the number of DoF for $\theta_T=0^o$ in Fig. \ref{fig_NumRes_1a} and $\theta_T=90^o$ in \ref{fig_NumRes_1b}, matches the heuristic number of DoF in th eparaxial setting, i.e.,  $m\approx \frac{L_T L_R}{\lambda d_0}$ \cite{Miller19}.
A second observation is that when the distance between the two arrays is reduced from $d_0=10\,m$ in Fig. \ref{fig_NumRes_1a} to $d_0 \approx 7\,m$ in Fig. \ref{fig_NumRes_1c}, the maximum number of DoF increases.
Another observation is the high number of modes available in the considered settings, for a small transmitting array and for a wide range of angles $\theta_T$ and $\theta_R$. An interesting conclusion based on the results obtained is that, in a multi-user scenario where the users are equipped with small arrays and the BS with a large one, the knowledge of the location and the relative orientation of the arrays may be used to design efficient resource allocation strategies, based on the number of DoF attainable at each small array.
\par
\begin{figure*}[!ht]
  \begin{subfigure}[t]{.33\linewidth}
  \includegraphics[trim=0 0 0 0,clip,width=\linewidth]{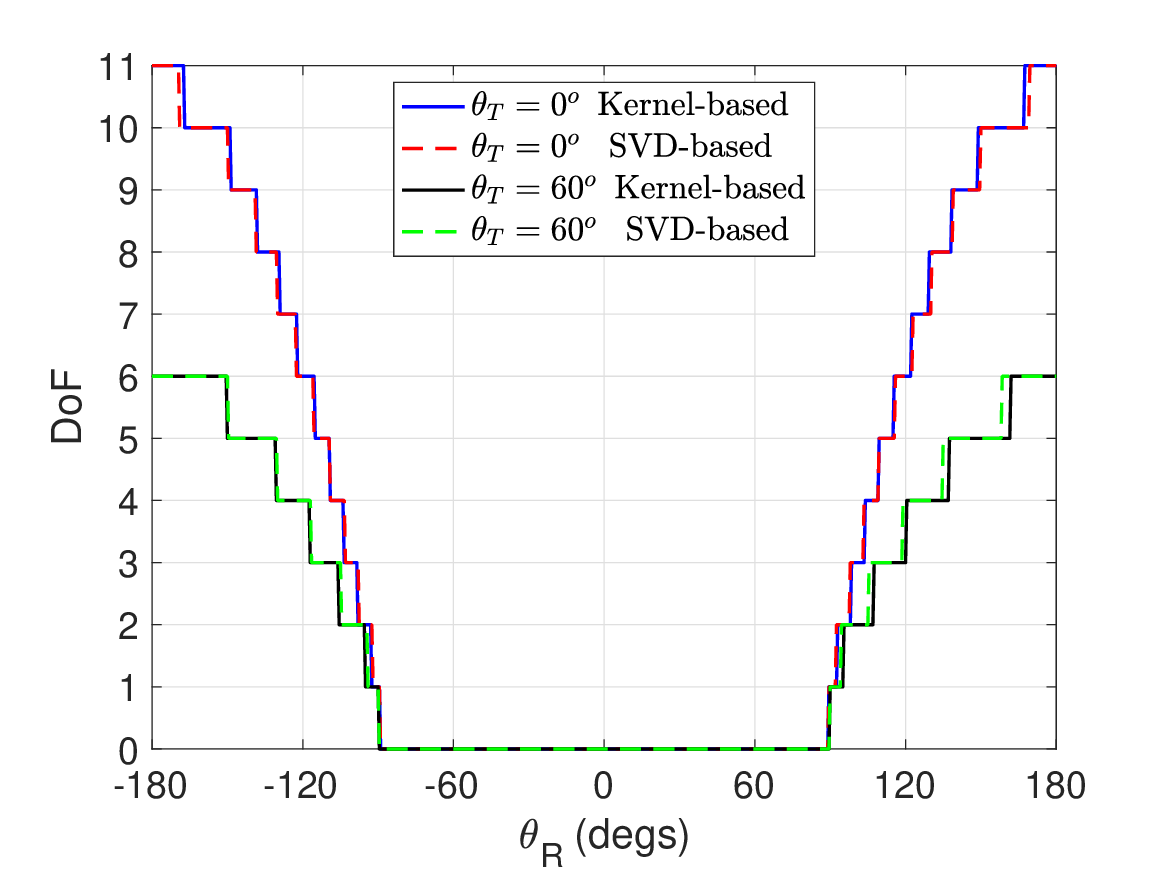}
   \caption{The DoF for $x_0 = 10m$, $y_0 = 0m$}
         \label{fig_Comp_m_SVD_a}
  \end{subfigure}\hfil
  \begin{subfigure}[t]{.33\linewidth}
    \includegraphics[trim=0 0 0 0,clip,width=\linewidth]{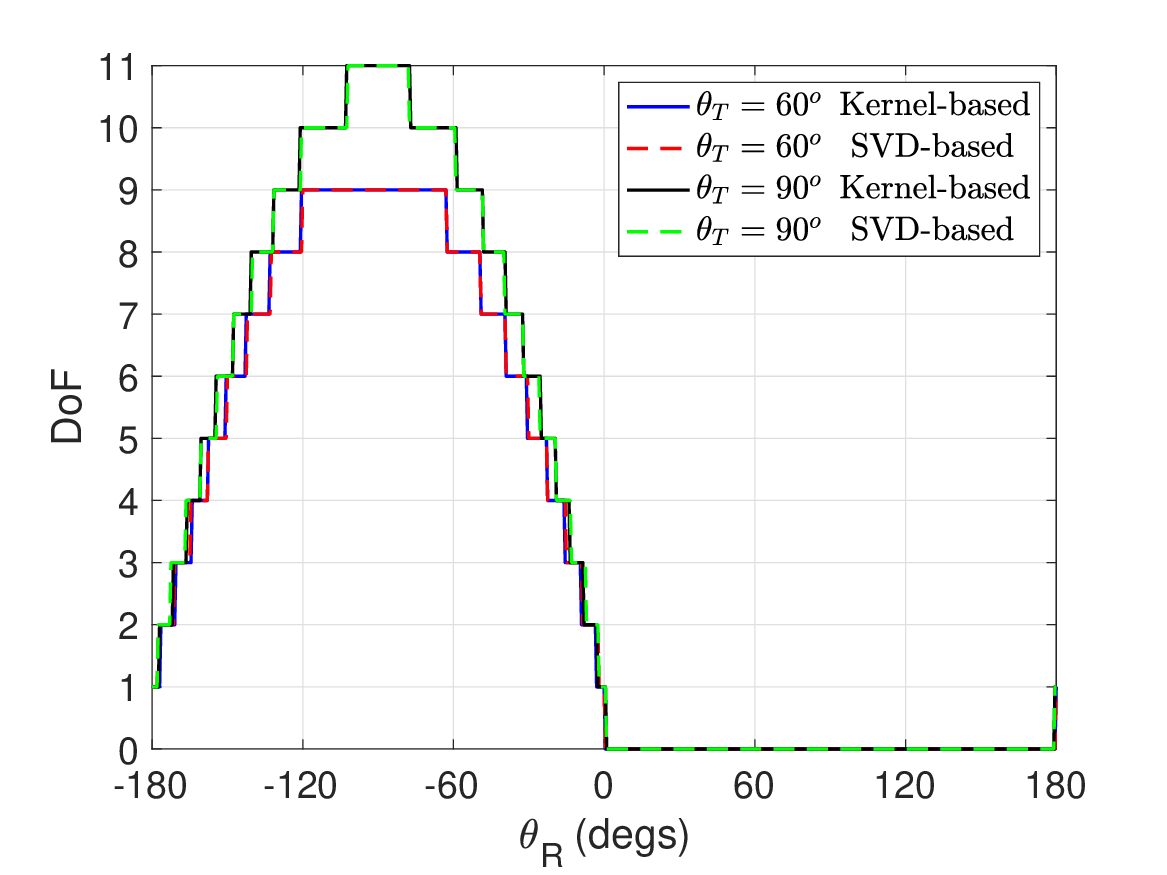}%
    \caption{The DoF for $x_0 = 0m$, $y_0 = 10m$}
         \label{fig_Comp_m_SVD_b}
  \end{subfigure}\hfil
  \begin{subfigure}[t]{.33\linewidth}
    \includegraphics[trim=0 0 0 0,clip,width=\linewidth]{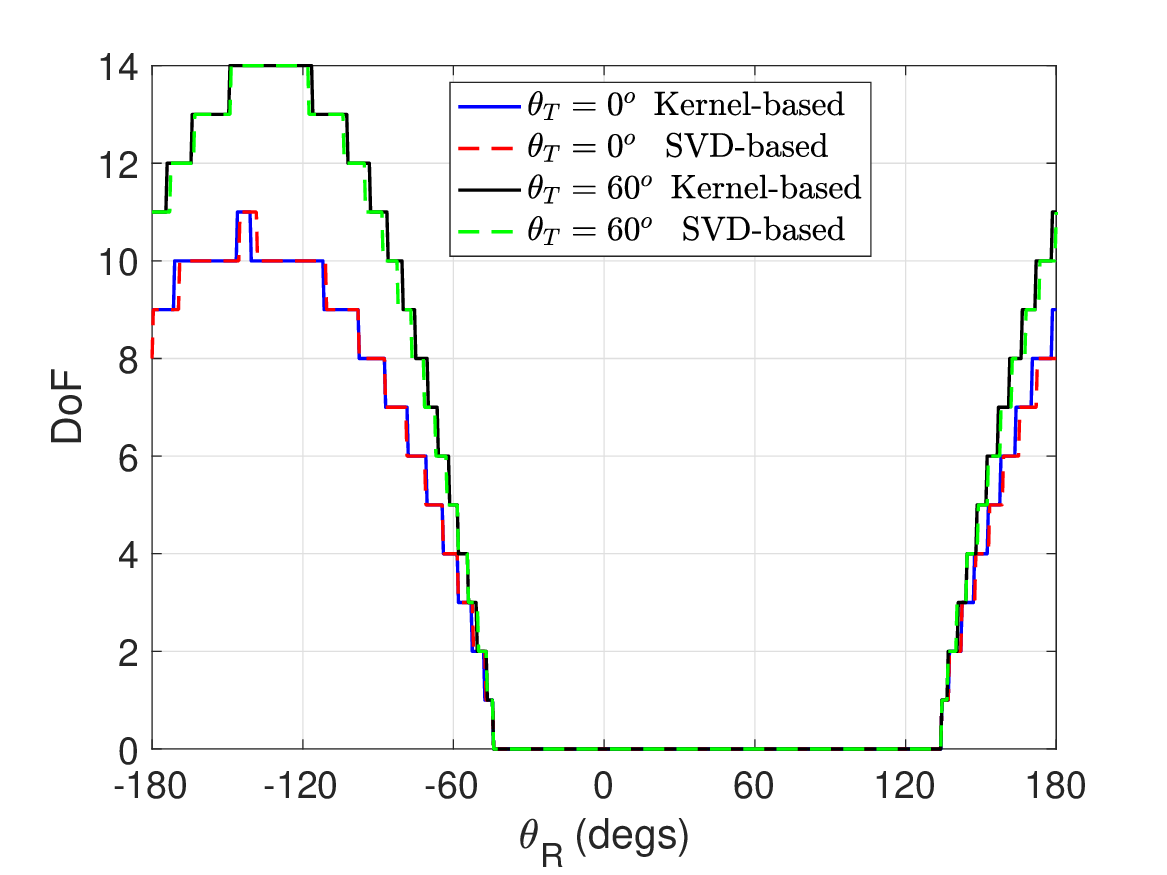}%
    \caption{The DoF for $x_0 = 5m$, $y_0 = 5m$}
         \label{fig_Comp_m_SVD_c}
  \end{subfigure}\hfil
   \caption{Comparison of the number of DoF calculated using the proposed kernel-based and the SVD-based method, for $L_T = 0.2m$, $L_R = 5m$, $f = 30\,$GHz.}\label{Comp_Kernel_SVD}
\end{figure*}

 A comparison of the number of DoF calculated by the proposed method and the SVD is shown in Fig. \ref{Comp_Kernel_SVD}. A round to the nearest integer has been applied to the kernel-based method for comparison with the SVD. The sub-figures correspond to two indicative values of $\theta_T$. The SVD was applied considering the same visibility conditions and effective length as calculated by Algorithm 1, and a threshold value 96\% of the sum-rule. The 11-th mode shown in Fig. \ref{fig_Comp_m_SVD_a} and \ref{fig_Comp_m_SVD_b} corresponds to a relatively weak eigenvalue.

\begin{figure}
\centering
  \includegraphics[width=0.75\linewidth]{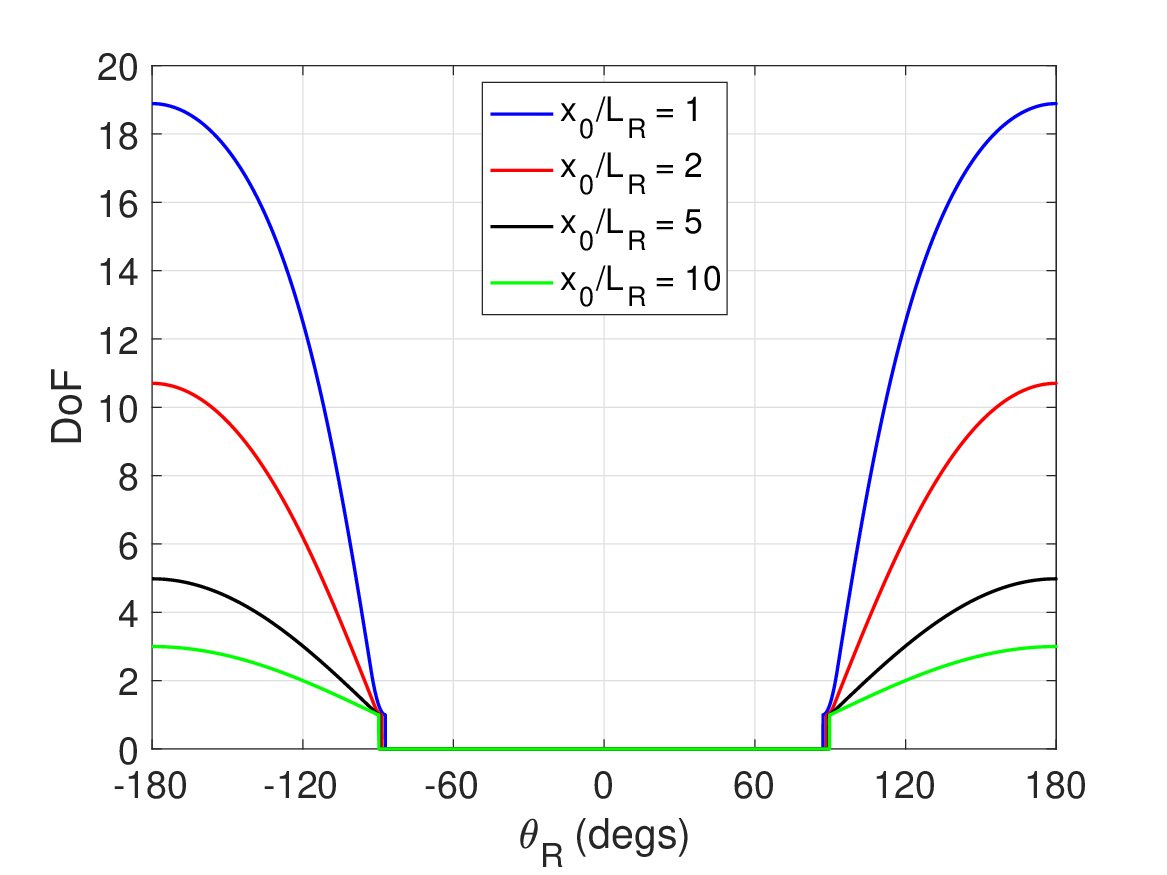}
  \caption{The DoF for $L_T = 0.2m$, $L_R = 2m$, $f = 30\,$GHz, $y_0 = 0m$, $\theta_T=0^o$, and different normalized distance $x_0/L_R$.}
  \label{fig_normx0}
\end{figure}

The impact of the distance between the two arrays on the number of DoF is depicted in Fig. \ref{fig_normx0}, where various normalized to $L_R$ distances are examined. This result reveals the advantage provided by near-field communications. Depending on the distance between the two arrays and the rotation angle $\theta_R$, the number of DoF takes values in the range [0,19], as expected from (\ref{DoF_case}). The zero value corresponds to the case where no visibility is achieved between the two arrays.

\subsection{Statistical Analysis}
\begin{figure*}[!ht]
\centering
  \begin{subfigure}[t]{.38\linewidth}
  \includegraphics[trim=0 0 0 0,clip,width=\linewidth]{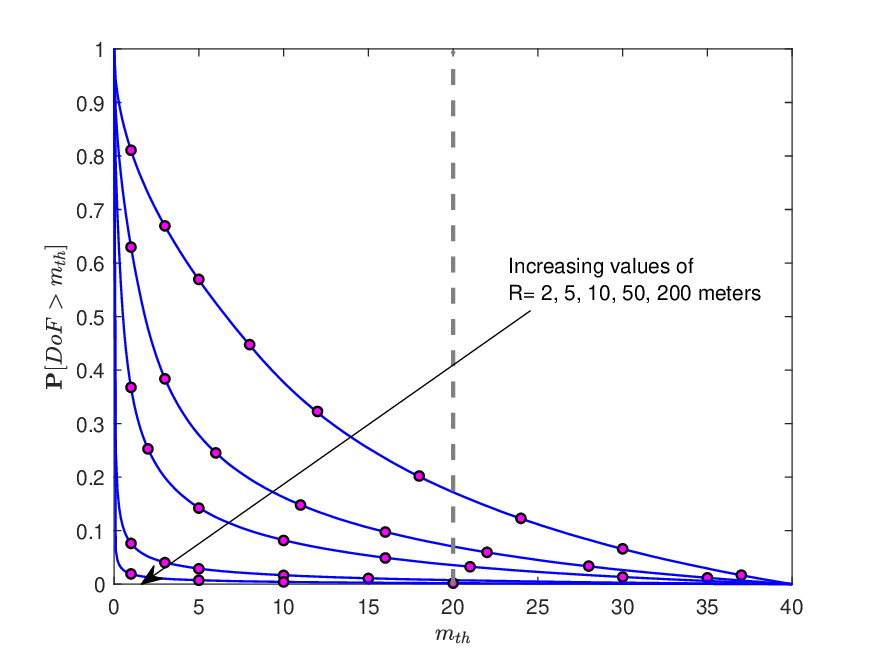}
   \caption{Partial visibility with $R^{+/-}$ being visible.}
         \label{fig_partial}
  \end{subfigure}\hfil
  \begin{subfigure}[t]{.38\linewidth}
    \includegraphics[trim=0 0 0 0,clip,width=\linewidth]{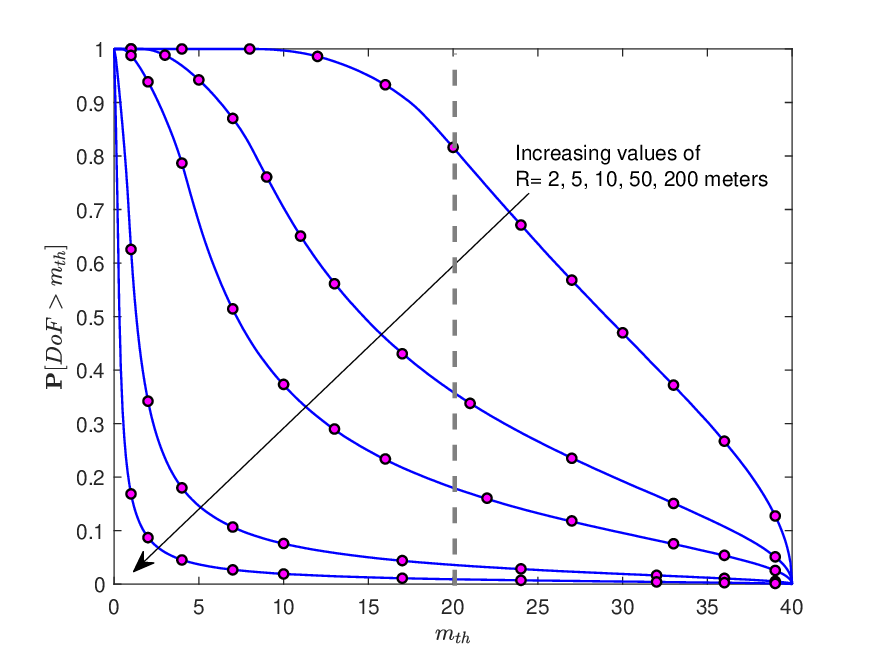}%
    \caption{Full visibility.}
         \label{fig_full}
  \end{subfigure}
   \caption{CCDF of the DoF versus $m_{th}$ for different values of the radius $R$ under partial and full visibility assumption and for $x_0$ being a random variable. Markers denote analytical results and the solid lines Monte Carlo simulations.}
   \label{fig_full_vs_partial}
\end{figure*}

In this subsection, the statistical behavior of the number of DoF is evaluated to gain system-level insights. The accuracy of the analytical results is verified against Monte-Carlo simulations. Unless stated otherwise, the following parameters are utilized: $f=30$ GHz, $L_T = 0.2$ meters, $L_R=2$ meters, and $R=20$ meters. Fig. \ref{fig_full_vs_partial} shows the complementary cumulative distribution function (CCDF), $F^c_{m}(m_{th})$, of the number of DoF for different values of $R$ and for $x_0$ being a random variable. The CCDF $F^c_{m}(m_{th})$ is defined as $F^c_{m}(m_{th})=1-\mathbb{P}[m\leq m_{th}]$ and is obtained through numerical integration of Theorem 1 and Theorem 2 under the partial visibility assumption and Theorem 3 under the full visibility assumption. A key observation is that a decrease in the range $R$ of $x_0$ significantly increases the probability of having a specific number of DoF. Therefore, as the distance $x_0$ decreases, the relative orientation of the arrays dominates the resulting DoF. Furthermore, the probability of achieving a target value of, e.g., $m_{th} = 20$ modes, which corresponds to approximately half of the maximum number of DoF, $2 L_T / \lambda$, dramatically increases with the decrease of $R$ in the full visibility scenario. This result highlights i) the need for establishing full visibility with the receiving array, and ii) the fact that the available number of DoF decreases rapidly with distance $x_0$. The latter is observed in Fig. \ref{fig_full}, where at a range of $R=200\,m$, i.e., well below the Fraunhofer distance $d_{FF} = 968\,m$, the probability of achieving $m>1$ is very low. Interestingly, the statistical behavior of the DoF is quite different between the partial and full visibility scenario for the smaller values of $R$ and the superiority in the number of DoF in full visibility is apparent. In fact, the probability that the number of DoF is greater than $m_{th} = 20$ in the full visibility scenario for $R=5$ meters is slightly lower than 40\%, while in the partial visibility scenario it is approximately 10\%. 

\begin{figure}[!t]
    \centering
    \includegraphics[keepaspectratio,width= 0.8\linewidth]{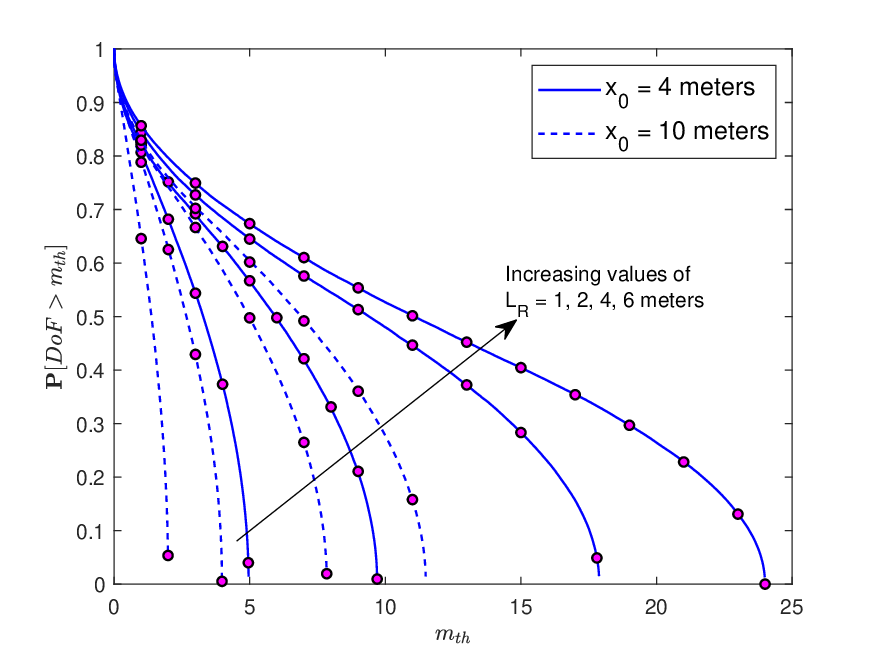}
       \caption{Conditional CCDF of the DoF versus $m_{th}$ for different values of $L_R$ and $x_0$. Markers denote analytical results and the lines Monte Carlo simulations.}
       \label{Conditional_CCDF}
\end{figure}

Fig. \ref{Conditional_CCDF} shows the conditional $F^c_{m}(m_{th})$ versus $m_{th}$ for two values of the coordinate $x_0$ and for different values of $L_R$.  The CCDF $F^c_{m}(m_{th})$ is obtained by numerical integration of Theorem 4. The first observation is that the maximum number of DoF is clearly location-dependent. Interestingly, a comb-like behavior is depicted for various values of $L_R$. Thus, the decrease in $F^c_{m}(m_{th})$ becomes steeper for larger receiving arrays and a higher number of DoF is more likely to be obtained. This is also a consequence of the higher probability that the receiving array is visible, which clearly depends on $x_0$ and $L_R$, as given by Definition 1. Fig. \ref{Conditional_CCDF} also shows that the increase in the number of DoF with the length of the Rx array $L_R$ is more profound for smaller values of $x_0$.

\begin{figure}[!t]
    \centering
    \includegraphics[keepaspectratio,width= 0.8\linewidth]{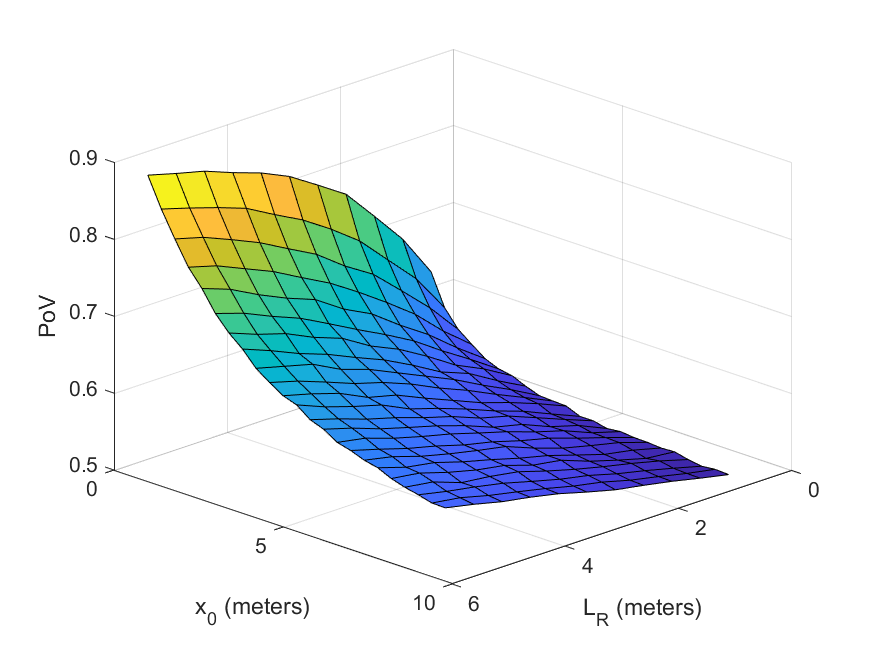}
       \caption{Probability of visibility versus $L_R$ and $x_0$.} \label{PoV}
\end{figure}

To elaborate more on system-level insights, Fig. \ref{PoV} depicts the PoV, $\mathcal{V}$, as given by Definition 1, versus $x_0$ and $L_R$. In particular, the increase in PoV is dominated by the close distance between the two arrays compared to an increase of $L_R$. Therefore, we conclude that \emph{a preferred array association policy should be primarily governed by the distance criterion and secondarily by the length of the arrays}.

\section{Conclusions}
In this work, the number of DoF between a small and a large linear array is examined, with an emphasis on the near-field. The proposed deterministic framework is applicable to general network deployments beyond the widely studied paraxial setting. In order to derive the number of DoF, the framework is based on the calculation of mutual visibility conditions between the two arrays. To this end, apart from the length of the two arrays, four more parameters are considered, namely the Cartesian coordinates of the center of the receiving array and two angles that model the rotation of each array around their center. The results are shown to coincide with those obtained with the SVD-based method. Subsequently, a stochastic geometry framework is proposed to capture the statistical behavior of the number of DoF. The analysis is performed for practical scenarios as a first step towards the investigation of the system-level performance in near-field communication networks. Among others, the results: i) quantified the advantage provided by near-field communications in mmWave networks, ii) highlighted the importance of the distance and relative orientation between the arrays, and iii) showed that, for the examined scenario, a preferred array association policy should be primarily based on the distance criterion and secondarily on the array length.
Two future research challenges have been identified: i) the investigation of the off-boresight near/far field boundaries for two large arrays in a wireless link, and ii) the extension of this work to planar arrays and an analysis in three dimensions.

\appendices
\appendices
\section{Proof of Lemma 1} 
The conditional PDF of $x_0$ assuming $x_0 > 0$ is given by integrating the joint PDF $f_{x_0,y_0} = 1/ \pi R^2$ over $y_0$ as 
\begin{equation}
f_{x_0}(x_0) = \frac{1}{\mathbb{P}[x_0>0]} \frac{2 \sqrt{R^2-x_0^2} }{\pi R^2}, 
\end{equation}
where $\mathbb{P}[x_0>0] = 1/2$ and $x_0 \in [0, R]$. Recalling \eqref{newalpha}, $\alpha_{-}$ can be simplified after applying some algebraic manipulations to $\alpha_{-}=\arctan\big(\frac{L_R}{2 x_0}\big)$. Conditioned on $x_0$, $\theta_T$ is uniformly and independently distributed in $[a_+^{max} - \frac{\pi}{2}, a_-^{max} - \frac{\pi}{2}]$ with the PDF given by $f_{\theta_T}(\theta) = \frac{1}{2  a_-^{max}}$, $\theta \in [a_+^{max} - \frac{\pi}{2}, a_-^{max} - \frac{\pi}{2}]$. Thus, $\alpha_{-}$ is no longer a random variable. Then, the conditional PDF of $A_- =\theta_T - a_-$ is given by
\begin{equation}\label{AppAfA_}
f_{A_{-}|x_0}(s) =  \frac{1}{2 \arctan \Big(\frac{L_R}{2 x_0}\Big)}, 
\end{equation}
for $s \in [-2\arctan\big(\frac{L_R}{2 x_0}\big)-\frac{\pi}{2},-\frac{\pi}{2}]$. After deconditioning over $x_0$ and solving $-2\arctan\big(\frac{L_R}{2 x_0}\big)-\frac{\pi}{2}=s$ for $s$ to make the ranges of $s$ independent of $x_0$, the PDF $f_{A_{-}}(s)$ is given by 
\begin{equation}
f_{A_{-}}(s) =  \int_{0}^{[x_{max}(\sin s)]^{-}} \frac{1}{\arctan \Big(\frac{L_R}{2 x_0}\Big)} \frac{2 \sqrt{R^2-x_0^2}}{\pi R^2}   {\rm{d}} x_0, 
\end{equation}
where $s \in [-\pi - \frac{\pi}{2}, - \frac{\pi}{2}]$. Finally, the PDF of $\rho_{-}$ is given through the change of variable $\rho_{-} = \sin A_{-} \Rightarrow A_{-} = \arcsin \rho_{-}$ by  
\begin{equation}
f_{\rho_{-}}(\rho) =  \frac{1}{\sqrt{1-\rho^2}} f_{A_{-}}(\arcsin \rho ), 
\end{equation}
for $\rho \in [-1, 1]$, which directly results in \eqref{Lemma1}. Next, recalling \eqref{newalpha}, $\alpha_{+}$ can be simplified to $\alpha_{+}=\arctan(-\cot \theta_T)$. The random variable $A_+ =\theta_T - a_+$ is written as
\begin{equation}
\begin{split}
& A_+ = \theta_T - a_+ =   \theta_T - \arctan(-\cot \theta_T)\\
& \myeqb -{\rm{arccot}} u - \arctan(-u) \myeqb  -\frac{\pi}{2},
\end{split}
\end{equation}   
where (a) follows through the transformation $\theta_T = {\rm{arccot}}(-u)$, (b) follows after exploiting the trigonometric identity $\arctan(-t)=-\frac{\pi}{2}+{\rm{arccot}}(t)$, which results in $A_+ = -\frac{\pi}{2}$. Finally, $\rho_{+}=\sin A_+ =-1$, and this completes the proof.

\section*{Acknowledgment}
This work was supported in part by the European Commission through the Horizon Europe project titled iSEE-6G under the GA No. 101139291, and the University of Piraeus Research Center (UPRC).
\\ H. S. Dhillon gratefully acknowledges the support of US NSF (Grants ECCS-2030215 and CNS-2107276). 
\\ The work of M. Di Renzo was supported in part by the European Union through the Horizon Europe project COVER under grant agreement number 101086228, the Horizon Europe project UNITE under grant agreement number 101129618, the Horizon Europe project INSTINCT under grant agreement number 101139161, and the Horizon Europe project TWIN6G under grant agreement number 101182794, as well as by the Agence Nationale de la Recherche (ANR) through the France 2030 project ANR-PEPR Networks of the Future under grant agreement NF-Founds 22-PEFT-0010, and by the CHIST-ERA project PASSIONATE under grant agreements CHIST-ERA-22-WAI-04 and ANR-23-CHR4-0003-01.


\begin{thebibliography}{10}

\bibitem{ICC2024}
A. G Kanatas, H. K. Armeniakos, and H. S. Dhillon, ``Degrees of Freedom With Small and Large Linear Surfaces in the Near Field”, \textit{IEEE Int. Conf. Commun.}, Denver, CO, USA, 2024, pp. 1932-1937. 

\bibitem{Huang20}
C. Huang, \emph{et al.,} ``Holographic MIMO Surfaces for 6G Wireless Networks: Opportunities, Challenges, and Trends,” \emph{IEEE Wireless Commun.}, vol. 27, no. 5, pp. 118–125, Oct. 2020.

\bibitem{Gong24}
T. Gong et al., ``Holographic MIMO Communications: Theoretical Foundations, Enabling Technologies, and Future Directions,"  \emph{IEEE Communications Surveys \& Tutorials,} vol. 26, no. 1, pp. 196-257, First quarter 2024.

\bibitem{Dardari21b}
D. Dardari and N. Decarli, ``Holographic Communication Using Intelligent Surfaces,” \emph{IEEE Commun. Mag. ,} vol. 59, no. 6, pp. 35-41, Jun. 2021.

\bibitem{Deng23}
R. Deng, Y. Zhang, H. Zhang, B. Di, H. Zhang, and L. Song, ``Reconﬁgurable holographic surface: A new paradigm to implement holographic radio,” \emph{IEEE Veh. Technol. Mag.}, vol. 18, no. 1, pp. 20-28, Mar. 2023.

\bibitem{diRenzo20}
M. Di Renzo et al., ``Smart Radio Environments Empowered by Reconfigurable Intelligent Surfaces: How It Works, State of Research, and The Road Ahead,” \emph{IEEE J. Sel. Areas Commun.,} vol. 38, no. 11, pp. 2450-2525, Nov. 2020.

\bibitem{Muharemovic08}
T. Muharemovic, A. Sabharwal, and B. Aazhang, ``Antenna packing in low-power systems: Communication limits and array design,” \emph{IEEE Trans. Inf. Theory,} vol. 54, no. 1, pp. 429–440, Jan. 2008.

\bibitem{Migliore08}
M. D. Migliore, ``On electromagnetics and information theory,” \emph{IEEE Trans. Antennas Propag.,} vol. 56, no. 10, pp. 3188–3200, Oct. 2008.

\bibitem{Zhu24}
J. Zhu, Z. Wan, L. Dai, M. Debbah, and H. V. Poor, ``Electromagnetic information theory: Fundamentals, modeling, applications, and open problems,” \emph{IEEE Wireless Commun.,} Early Access, 2024.

\bibitem{France18}
M. Franceschetti, ``Wave Theory of Information”, Cambridge Univ. Press, 2018.

\bibitem{Pizzo20}
A. Pizzo, T. L. Marzetta and L. Sanguinetti, ``Degrees of Freedom of Holographic MIMO Channels,” 2020 IEEE 21st International Workshop on Signal Processing Advances in Wireless Communications (SPAWC), Atlanta, GA, USA, 2020, pp. 1-5.

\bibitem{Dardari20}
D. Dardari, ``Communicating With Large Intelligent Surfaces: Fundamental Limits and Models”, \emph{IEEE J. Sel. Areas Commun.,} vol. 38, no. 11, pp. 2526-2537, Nov. 2020.

\bibitem{Marco24}
M. D. Renzo and M. D. Migliore, ``Electromagnetic Signal and Information Theory," \emph{IEEE BITS the Info Theory Mag.,} doi: 10.1109/MBITS.2024.3359523.

\bibitem{Sayeed02}
A. M. Sayeed, ``Deconstructing multiantenna fading channels," \emph{IEEE Trans. on Signal Processing,} vol. 50, no. 10, pp. 2563-2579, Oct. 2002.

\bibitem{Tse05}
D. Tse and P. Viswanath, \emph{Fundamentals of Wireless Communication.} Cambridge, U.K.: Cambridge Univ. Press, 2005.

\bibitem{Kalis08}
A. Kalis, A. G. Kanatas and C. B. Papadias, ``A Novel Approach to MIMO Transmission Using a Single RF Front End," \emph{IEEE J. Sel. Areas Commun.,} vol. 26, no. 6, pp. 972-980, August 2008.

\bibitem{Barousis11}
V. I. Barousis, A. G. Kanatas and A. Kalis, ``Beamspace-Domain Analysis of Single-RF Front-End MIMO Systems," \emph{IEEE Trans. Veh. Technol.,} vol. 60, no. 3, pp. 1195-1199, March 2011.

\bibitem{Brady13}
J. Brady, N. Behdad and A. M. Sayeed, ``Beamspace MIMO for Millimeter-Wave Communications: System Architecture, Modeling, Analysis, and Measurements," \emph{IEEE Trans. on Antennas and Propag.,} vol. 61, no. 7, pp. 3814-3827, July 2013.

\bibitem{Vasileiou13}
P. N. Vasileiou, K. Maliatsos, E. D. Thomatos and A. G. Kanatas, ``Reconfigurable Orthonormal Basis Patterns Using ESPAR Antennas," \emph{IEEE Antennas and Wireless Propag. Lett.,} vol. 12, pp. 448-451, 2013.

\bibitem{Lin20}
S. Lin, Z. Peng and T. M. Antonsen, "A Stochastic Green’s Function for Solution of Wave Propagation in Wave-Chaotic Environments," in IEEE Transactions on Antennas and Propagation, vol. 68, no. 5, pp. 3919-3933, May 2020.

\bibitem{PizzoSang22}
A. Pizzo, L. Sanguinetti and T. L. Marzetta, "Spatial Characterization of Electromagnetic Random Channels," in IEEE Open Journal of the Communications Society, vol. 3, pp. 847-866, 2022.

\bibitem{Liu23}
Y. Liu, Z. Wang, J. Xu, C. Ouyang, X. Mu and R. Schober, ``Near-Field Communications: A Tutorial Review” \emph{IEEE Open J. of the Commun. Soc.,} vol. 4, pp. 1999-2049, 2023.

\bibitem{Cui23}
M. Cui, Z. Wu, Y. Lu, X. Wei, and L. Dai, ``Near-ﬁeld MIMO communications for 6G: Fundamentals, challenges, potentials, and future directions,” \emph{IEEE Commun. Mag.,} vol. 61, no. 1, pp. 40–46, Jan. 2023.

\bibitem{Lu22}
H. Lu and Y. Zeng, ``Communicating With Extremely Large-Scale Array/Surface: Unified Modeling and Performance Analysis," \emph{IEEE Trans. Wireless Commun.,} vol. 21, no. 6, pp. 4039-4053, June 2022.

\bibitem{Zhang23}
H. Zhang, N. Shlezinger, F. Guidi, D. Dardari and Y. C. Eldar, ``6G Wireless Communications: From Far-Field Beam Steering to Near-Field Beam Focusing,” \emph{IEEE Commun. Mag.,} vol. 61, no. 4, pp. 72-77, Apr. 2023.

\bibitem{Bjornson21}
E. Bj\"{o}rnson, \"{O}. T. Demir, and L. Sanguinetti, ``A primer on nearﬁeld beamforming for arrays and reconﬁgurable intelligent surfaces,” \emph{Proc. 55th Asilomar Conf. Signals, Syst., Comput,} Oct. 2021, pp. 105–112.

\bibitem{Myers22}
N. J. Myers and R. W. Heath, ``Infocus: A spatial coding technique to mitigate misfocus in near-ﬁeld LoS beamforming,” \emph{IEEE Trans. Wireless Commun.,} vol. 21, no. 4, pp. 2193–2209, Apr. 2022.

\bibitem{Hu18}
S. Hu, F. Rusek, and O. Edfors, ``Beyond massive MIMO: The potential of data transmission with large intelligent surfaces,” \emph{IEEE Trans. Signal Process.,} vol. 66, no. 10, pp. 2746–2758, May 2018.

\bibitem{Pizzo22}
A. Pizzo, A. D. J. Torres, L. Sanguinetti, and T. L. Marzetta, ``Nyquist sampling and degrees of freedom of electromagnetic ﬁelds,” \emph{IEEE Trans. Signal Process.,} vol. 70, pp. 3935–3947, 2022.

\bibitem{Ji23}
R. Ji et al., ``Extra DoF of Near-Field Holographic MIMO Communications Leveraging Evanescent Waves," \emph{IEEE Wireless Commun. Lett.,} vol. 12, no. 4, pp. 580-584, April 2023.

\bibitem{Miller00}
D. A. B. Miller, ``Communicating with waves between volumes: Evaluating orthogonal spatial channels and limits on coupling strengths”, \emph{Appl. Opt.,} vol. 39, no. 11, pp. 1681-1699, Apr. 2000.

\bibitem{Do23}
H. Do, N. Lee and A. Lozano, "Parabolic Wavefront Model for Line-of-Sight MIMO Channels," in IEEE Transactions on Wireless Communications, vol. 22, no. 11, pp. 7620-7634, Nov. 2023.

\bibitem{Pu15}
X. Pu, S. Shao, K. Deng and Y. Tang, ``Effects of Array Orientations on Degrees of Freedom for 3D LoS Channels in Short-Range Communications," \emph{IEEE Wireless Commun. Lett.,} vol. 4, no. 1, pp. 106-109, Feb. 2015.

\bibitem{Bucci87}
O. Bucci and G. Franceschetti, ``On the spatial bandwidth of scattered fields," \emph{IEEE Trans. on Antennas and Propag.,} vol. 35, no. 12, pp. 1445-1455, Dec. 1987.

\bibitem{Bucci89}
O. M. Bucci and G. Franceschetti, ``On the degrees of freedom of scattered fields,'' \emph{IEEE Trans. on Antennas and Propag.,} vol. 37, no. 7, pp. 918-926, July 1989.

\bibitem{Bucci98}
O. M. Bucci, C. Gennarelli and C. Savarese, ``Representation of electromagnetic fields over arbitrary surfaces by a finite and nonredundant number of samples," \emph{IEEE Trans. on Antennas and Propag.,} vol. 46, no. 3, pp. 351-359, March 1998.

\bibitem{Franceschetti11}
M. Franceschetti, M. D. Migliore, P. Minero and F. Schettino, ``The Degrees of Freedom of Wireless NetworksVia Cut-Set Integrals," \emph{IEEE Trans. Inf. Theory,} vol. 57, no. 5, pp. 3067-3079, May 2011.

\bibitem{Landau75}
H. J. Landau, ``On Szegö’s eingenvalue distribution theorem and non-Hermitian kernels," \emph{Journal d’Analyse Mathématique,} vol. 28, no. 1, pp. 335–357, 1975.

\bibitem{Pizzo22b}
A. Pizzo and A. Lozano, ``On Landau’s Eigenvalue Theorem for Line-of-Sight MIMO Channels," \emph{IEEE Wireless Comm. Lett.,} vol. 11, no. 12, pp. 2565-2569, Dec. 2022.

\bibitem{Ruiz23}
J.C. Ruiz-Sicilia et al., ``On the Degrees of Freedom and Eigenfunctions of Line-of-Sight Holographic MIMO Communications," 2023. [Online]. Available:https://arxiv.org/abs/2308.08009.

\bibitem{Chen24}
H. Chen, S. Yue, M. Di Renzo, and H. Zhang, "Degrees of Freedom of Holographic MIMO in Multi-user Near-field Channels," 2025. [Online]. Available:https://arxiv.org/abs/2410.05013.

\bibitem{Marco23}
M. Di Renzo, D. Dardari and N. Decarli, ``LoS MIMO-Arrays vs. LoS MIMO-Surfaces," in \emph{17th European Conference on Antennas and Propagation (EuCAP)}, Florence, Italy, 2023, pp. 1-5.

\bibitem{Dardari21a}
N. Decarli and D. Dardari, ``Communication Modes With Large Intelligent Surfaces in the Near Field,” \emph{IEEE Access,} vol. 9, pp. 165648-165666, 2021.

\bibitem{Ding22}
L. Ding, E. G. Ström and J. Zhang, ``Degrees of Freedom in 3D Linear Large-Scale Antenna Array Communications—A Spatial Bandwidth Approach," \emph{IEEE J. Sel. Areas Commun.,} vol. 40, no. 10, pp. 2805-2822, Oct. 2022.

\bibitem{Ding24}
L. Ding, J. Zhang and E. G. Ström, ``Spatial Bandwidth Asymptotic Analysis for 3D Large-Scale Antenna Array Communications," \emph{IEEE Trans. Wireless Commun.,}, vol. 23, no. 4, pp. 2638-2652, April 2024.

\bibitem{Schober}
Y. Liu, Z. Wang, J. Xu, C. Ouyang, X. Mu and R. Schober, ``Near-Field Communications: A Tutorial Review," \emph{ IEEE Open J. Commun. Soc.}, vol. 4, pp. 1999-2049, Aug. 2023.

\bibitem{Monemi}
M. Monemi, S. Bahrami, M. Rasti and M. Latva-aho, "A Study on Characterization of Near-Field Sub-Regions for Phased-Array Antennas," \emph{IEEE Trans. Commun.,} vol. 73, no. 5, pp. 2964-2979, May 2025.

\bibitem{Hughes}
Hughes, J., Van Dam, A. Mcguire, M., Sklar, D., Foley, J., Feiner, S., and Akeley, K., \emph{Computer Graphics: Principles and Practice,} 3rd edition, Pearson Education Inc., 2014. pg.159.

\bibitem{Poon05}
A. S. Y. Poon, R. W. Brodersen, and D. N. C. Tse, “Degrees of freedom in multiple-antenna channels: A signal space approach,” \emph{IEEE Trans. Inf. Theory,} vol. 51, no. 2, pp. 523-536, Feb. 2005.

\bibitem{Sherman62}
J. Sherman, "Properties of focused apertures in the Fresnel region,", \emph{IRE Trans. on Antennas and Propag.}, vol. 10, no. 4, pp.399-408, July 1962.

\bibitem{Ryzhik}
I. S. Gradshteyn and I. M. Ryzhik (Eds.). \emph{{T}able of {I}ntegrals, {S}eries, and {P}roducts,} 8th edition. Academic Press, 2015.

\bibitem{Miller19}
D.A.B. Miller, "Waves, modes, communications, and optics: a tutorial," \emph{Adv. Opt. Photon.} vol.11, pp.679-825, 2019.

\end{thebibliography}
\end{document}